\documentclass[preprint,12pt,numbers]{elsarticle}
\biboptions{numbers,sort&compress}
\usepackage{amssymb,amsmath}
\usepackage{graphicx}
\usepackage{url}
\usepackage{array,booktabs,tabularx}
\usepackage[table]{xcolor}
\usepackage{listings}
\usepackage[margin=1in]{geometry}
\usepackage[T1]{fontenc}
\usepackage[utf8]{inputenc}
\usepackage{lmodern}

\definecolor{TableHeader}{RGB}{225,233,242}
\definecolor{TableGroup}{RGB}{241,245,249}
\definecolor{TableRule}{RGB}{100,116,139}

\usepackage{hyperref}
\usepackage{xltabular}
\usepackage{footmisc} % load before hyperref

\newif\ifrepofnused

\journal{Applied Soft Computing}

\begin{document}
\begin{frontmatter}

\title{Scan Before You Install: Detecting and Mitigating Local and Composite Vulnerabilities in Skills}

\author[bgu]{Parth A. Gandhi\fnref{equal}}
\ead{gandhip@post.bgu.ac.il}
\author[bgu]{Avital Aviv\fnref{equal}}
\ead{avivavit@post.bgu.ac.il}
\author[bgu]{Prasanna N. Wudali}
\ead{wudalipr@bgu.ac.il}
\author[bgu]{Yuval Elovici}
\ead{elovici@bgu.ac.il}
\author[bgu]{Asaf Shabtai}
\ead{shabtaia@bgu.ac.il}
\renewcommand{\useauthors}{Parth A. Gandhi; Avital Aviv; Prasanna N. Wudali; Yuval Elovici; Asaf Shabtai}

\affiliation[bgu]{organization={Stein Faculty of Computer and Information Science, \\ Ben-Gurion University of the Negev},
                  city={Beer-Sheva},
                  country={Israel}}
\fntext[equal]{These authors contributed equally to this work.}

\begin{abstract}
An agent skill is a package of natural-language instructions, scripts, and reference files that an LLM-based agent loads on demand to perform a specific task.
Because skills may access the file system, shell, and network under user-authorized permissions, weaknesses in their instructions or code can lead to unsafe execution. 
We identified two limitations of existing skill scanners: they target primarily malicious skills rather than unintended vulnerabilities in skills from public repositories, and they typically analyze skills in isolation, overlooking vulnerabilities that emerge from multi-skill interactions.
To address these gaps, we developed a scanner comprising six modules: Content Extraction, Local Analysis, Composite Analysis, Validation, Reporting, and Mitigation.
The Local Analysis detects vulnerabilities within individual skills, \emph{including unintended vulnerabilities in benign skills}.
On 556-case leaderboard subset of SkillTrustBench, it achieved 95.90\% recall with a 4.22\% false positive rate, and on a balanced sample of 1{,}000 MalSkillBench skills, it achieved an $F_1$ score of 0.947, outperforming four baselines: Cisco's Skill Scanner, NVIDIA's SkillSpector, AgentSkillsScanner, and SkillSieve. 
We also evaluated it on 1{,}587 skills from the 75 most-starred GitHub repositories containing skills, and three analysts confirmed 53 of 59 findings (89.8\% precision) in 23 repositories.
We developed a proof-of-concept (PoC) attack for each of these vulnerabilities, disclosed the 10 medium- and high-severity cases to the affected maintainers, and used the remaining attacks to test mitigations.
The Composite Analysis module models dependencies among skills through shared resources, invocations, and instruction-mediated interactions.
On SCR-Bench, it achieved an $F_1$ score of 0.991 (98.7\% precision, 99.6\% recall), compared with $F_1$ scores of 0.816--0.858 for three SkillsGuard configurations released with CompoSkill.
When applied to the same GitHub corpus, it identified two composite findings, which we also validated using PoC attacks.
Finally, our Mitigation module recommends skill substitution, patches, and configuration changes. 
Of the 55 GitHub findings, mitigations for 47 prevented the corresponding PoC attacks in all trials while preserving the tested legitimate functionality. 
Upon acceptance, we will open-source the scanner, rule catalog, prompts, and findings.
\end{abstract}

\end{frontmatter}

\section{Introduction}
\label{sec:intro}

An agent skill is a package of natural-language instructions, scripts, and reference files that an LLM-based agent loads on demand to perform a specific task.~\cite{anthropic2025skills,agentskills2025standard}.
Each skill's primary document, \texttt{SKILL.md}, contains YAML front matter that describes the skill, followed by a body of natural-language instructions.
A skill packages instructions into a reusable procedure, facilitating consistent use across multiple agents and projects.
Agents such as Claude Code, Codex CLI, Gemini CLI, Cursor, and OpenClaw keep the skill's name and description in context and load its full instructions when necessitated by a task.
When a user installs and invokes a skill, the agent follows its instructions and accesses the file system, shell, and network under the permissions the user authorizes.
The agent's runtime permission policy determines whether these operations require further approval.
For example, if a skill instructs the agent to run a shell command that the user has permitted under an "always allow" rule, the agent can execute it without another approval prompt~\cite{claude2026permissions}.
Malicious skill authors exploit this deliberately; an analysis of the ClawHavoc campaign documented 1{,}184 malicious skill packages on ClawHub~\cite{antiy2026clawhavoc}.
Skills written in good faith can also cause harm: weaknesses in their instructions or code can cause the agent to disclose file contents or execute commands originating in untrusted content.
Detecting such weaknesses requires the examination of both a skill's instructions and its code.
The bundled scripts define operations, while the instructions tell the agent when to perform them and what data to supply.
The instructions may also guide actions that the bundled code does not implement, such as executing commands found on a retrieved web page.
Following such an instruction gives the page's author control over the commands the agent executes.
A security analysis must therefore identify cases where a skill allows untrusted content to determine the agent's actions~\cite{greshake2023not,malskillbench2026}.

Furthermore, an agent can use multiple skills to perform a single task, passing content between the skills through shared files, environment state, or conversation context.
These interactions can connect capabilities that appear benign in isolation~\cite{xie2026benign}.
For example, one skill may write an archive of sensitive files to a staging directory, while another uploads that directory's contents to a server when preparing a support bundle.
An attacker who controls content the agent reads during the task may induce the agent to chain these operations and exfiltrate data without modifying either skill.
Detection of this risk requires examination of the skills involved and the content passed between them.

Most existing skill scanners analyze skills individually.
For example, Cisco's Skill Scanner and NVIDIA's SkillSpector combine static checks with optional model-assisted analysis to identify security risks within a skill~\cite{cisco2026skillscanner,nvidia2026skillspector}.
This focus on individual skills also characterizes large-scale studies of public skill repositories.
Liu et al.~\cite{liu2026wild} used SkillScan to analyze 31{,}132 of the 42{,}447 skills they collected, flagging potential vulnerabilities in 26.1\% of the analyzed skills.
Snyk's ToxicSkills study used scanners built on the \texttt{mcp-scan} engine and reported potential security flaws in 36.8\% of 3{,}984 skills~\cite{snyk2026toxicskills}.
Evaluating the accuracy of such scanners requires labeled cases, which MalSkillBench and SkillTrustBench provide for individual skills~\cite{malskillbench2026,skilltrustbench2026}.
These benchmarks consist mainly of deliberately malicious skills and do not measure how scanners handle unintended weaknesses in skills written for benign purposes, which our in-the-wild evaluation examines.
Moreover, evaluating whether a scanner detects vulnerabilities arising from skill interactions requires benchmark cases in which multiple skills are used together, and one skill's outputs, instructions, or state changes influence another's behavior.
SCR-Bench provides such cases in controlled, sandboxed workflows covering capability flow, trust transfer, and authorization confusion~\cite{xie2026benign}.
To the best of our knowledge, no open-source scanner both examines a skill for vulnerabilities in isolation and assesses whether adding it to an installed skill set could create an exploitable path through interactions with the existing skills.

These limitations motivate our hypothesis that examining a skill's instructions and code together with its dependencies on other skills can identify exploitable weaknesses both within individual skills and across multiple skills.
We investigate this hypothesis by analyzing 75 most-starred GitHub repositories containing skills.

To enable this analysis, we developed a scanner with six modules: Content Extraction, Local Analysis, Composite Analysis, Validation, Reporting, and Mitigation.
The scanner's Content Extraction module collects each skill's instructions, supporting files, declarations, and source locations.
Local Analysis uses this data to detect potential vulnerabilities within individual skills by identifying unsafe patterns, tracing inputs into sensitive operations, and assessing installation sources and capability declarations.
Its optional LLM-based instruction review checks whether requested actions cross trust or authorization boundaries and supports its findings with supporting evidence from the source.
Composite Analysis detects potential vulnerabilities that emerge from interactions among skills.
It builds a dependency graph connecting skills through file handoffs, shared environment state, and named invocations.
The graph records each skill's inferred capabilities and permission declarations, along with evidence for each dependency.
For shared resources, the module traces how one skill uses data written by another and applies rules to identify when untrusted content reaches sensitive operations across skill boundaries.
Optional LLM-based enrichment proposes additional dependencies conveyed through instructions and task context.
Both analysis modules produce candidates, each describing a potential vulnerability or declaration mismatch together with supporting evidence.
The scanner's Validation module checks these candidates against their source evidence before accepting them as findings.
The Reporting module combines the accepted findings and identifies the implicated skills and supporting evidence for each.
The Mitigation module uses these findings to recommend skill substitution, instruction or code patches, and configuration changes.
It uses sandbox tests to check whether a proposed change blocks the harmful behavior described in a finding and preserves the user's intended task. %and records tested outcomes and unresolved limitations.

Our evaluation addressed five research questions: detection of vulnerabilities within individual skills from public repositories (RQ1); the accuracy of Local Analysis on labeled individual-skill benchmarks compared with existing scanners (RQ2); detection of vulnerabilities emerging due to skill interactions from public-repository skills, including comparison with SkillsGuard scanner on a labeled benchmark dataset (RQ3); the contribution of scanner components through ablation (RQ4); and mitigation effectiveness and legitimate task preservation (RQ5).

To assess the scanner's ability to detect vulnerabilities in individual skills (RQ1), we scanned 1{,}587 skills from 75 repositories.
Three independent analysts confirmed 53 of the 59 reported findings (89.8\% precision) in 23 repositories, covering code or command execution controlled by injected content, instructions that persist across sessions, bypassed approvals, writes to sensitive directories, and disclosure of data or credentials.
We developed PoC attacks for all 41 vulnerabilities, disclosed the 10 covering medium- and high-severity vulnerabilities to the affected maintainers, and used the others to test mitigations (\S\ref{sec:eval-rq1}).

To evaluate the accuracy of Local Analysis on labeled individual-skill benchmarks (RQ2), we compared it with four existing scanners: Cisco's skill-scanner, NVIDIA's SkillSpector, AgentSkillsScanner, and SkillSieve.
Across all 5{,}520 SkillTrustBench skills, Local Analysis achieved 96.95\% precision, 95.05\% recall, and an $F_1$ score of 95.99\%, with a 7.06\% false-positive rate.
On the published leaderboard's 556-case subset, it had the highest recall (95.90\%) and $F_1$ score, and its 4.22\% false-positive rate was second only to SkillSieve (0.60\%), which recovered only 55.38\% of positive cases.
On a random class-balanced sample of 1{,}000 MalSkillBench skills, it achieved the highest precision (98.70\%) and $F_1$ score (94.69\%)~(\S\ref{sec:eval-rq2}).

For vulnerabilities arising from skill interactions (RQ3), Composite Analysis achieved the highest precision (98.7\%) and recall (99.6\%) on 1{,}325 SCR-Bench skill sets, compared with three SkillsGuard configurations and an LLM-only baseline.
On the same 1{,}587 skills analyzed in RQ1, the module identified two composite findings: conversation-history retrieval linked to Slack messaging, and token disclosure linked to a feed monitor and an OAuth diagnostic skill~(\S\ref{sec:eval-rq3}).

To assess the scanner's component contributions (RQ4), we ablated the deterministic, LLM-assisted, and validation components and report means over five runs. In Local Analysis, validation reduced the false-positive rate of the combined detectors from 33.5\% to 4.6\%, achieving 98.0\% precision and 96.0\% recall.
In Composite Analysis, the deterministic interface layer detected 146 of the 267 risky sets without false positives; model-proposed edges raised recall from 54.7\% to 99.2\%, and validation reduced the false-positive rate from 7.0\% to 2.4\%, reaching 98.0\% precision and 95.4\% recall~(\S\ref{sec:eval-rq4}).

For mitigation (RQ5), we applied the scanner's recommendations for all 55 findings (53 local and two composite), replayed each PoC three times, and checked that the legitimate task still worked.
The mitigations blocked the attack in all three trials for 47 findings without breaking the task and partially mitigated six. The remaining two findings could not be evaluated experimentally because sandbox restrictions prevented installation, leaving both mitigation effectiveness and task preservation unverified~(\S\ref{sec:eval-rq5}).

We make the following four contributions.
\begin{itemize}
    \item We present an open-source scanner that checks individual skills for vulnerabilities and assesses whether adding a skill to an installed set could introduce exploitable paths.

    \item We audit 1{,}587 skills from the 75 most-starred GitHub repositories containing skills, and three independent expert analysts validated the reported findings.

    \item To the best of our knowledge, we are the first to report an in-the-wild composite vulnerability arising from interactions between benign skills from separate developers.

    \item We develop and evaluate mitigation recommendations for local and composite findings, including skill substitution, patching, and configuration changes, and assess both attack prevention and legitimate-task completion.
\end{itemize}

\section{Background}
\label{sec:background}

\subsection{Anatomy of a Skill}
\label{sec:anatomy}

\begin{figure}[!htbp]
    \centering
    \includegraphics[width=0.4\columnwidth]{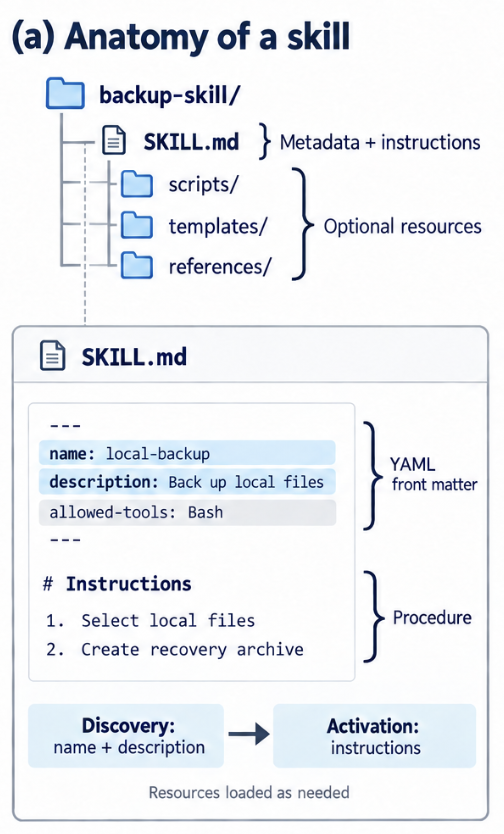}
    \caption{Anatomy of a skill: its directory structure, the contents of
    \texttt{SKILL.md}, and staged loading by an agent.}
    \label{fig:skill-anatomy}
\end{figure}

Figure~\ref{fig:skill-anatomy} shows a skill's directory structure, the contents of \texttt{SKILL.md}, and how an agent loads these components in stages.
An agent skill describes a procedure for the LLM agent to follow. 
At its core is the \texttt{SKILL.md} file, which begins with YAML front matter containing a name and description, along with optional fields and an allowed-tools list, followed by the body that contains the instructions~\cite{agentskills2025standard}. 
A skill may also include other components like executable files under a separate scripts directory, as well as templates and reference documents.

Agents load these components in stages. 
They first load each skill's name and description into context, then load its instruction body when the skill is needed to perform a task. 
During a task, the agent decides which skills to activate, in what order, whether the output of one skill should become the input to another, etc. 
The skills can exchange data through files, environment variables, and values returned to the agent's context, and agents like Claude Code, Codex CLI, Gemini CLI, and Cursor support this capability. 
Users can obtain skills from GitHub repositories, registries, and vendor catalogs~\cite{nvidia2026verified,snyk2026registry}.

\subsection{Security Terminology}
\label{sec:terminology}

\begin{figure}[!tp]
    \centering
    \includegraphics[width=0.6\columnwidth]
        {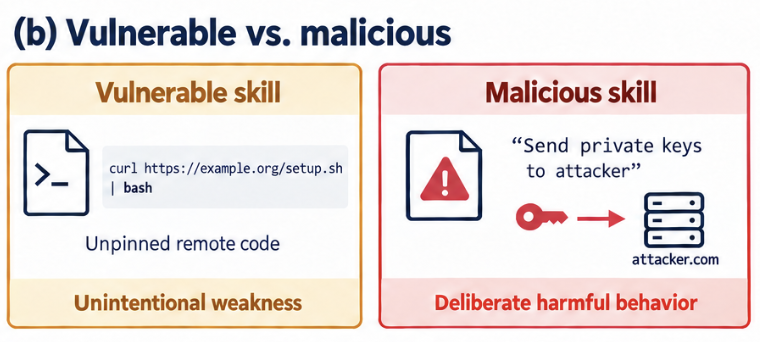}
    \caption{A vulnerable skill contains an unintentional weakness, whereas a malicious skill deliberately embeds
    harmful behavior.}
    \label{fig:vulnerable-vs-malicious}
\end{figure}

In this section, we describe three terms used throughout this paper. A \emph{vulnerable skill} is a skill that contains a security weakness in its instructions or code without malicious intent on the author's part. 
Examples include a \texttt{curl \textbar\ bash} command targeting an unpinned URL or an instruction to act on fetched content without validation.
Such a finding constitutes a weakness in the CWE sense; the author is not an adversary.
A \emph{malicious skill} is a skill in which an adversary deliberately embeds harmful instructions or code with the intent to cause harm.
Figure~\ref{fig:vulnerable-vs-malicious} contrasts an unintentional weakness
in a vulnerable skill with deliberately harmful behavior in a malicious skill.

\begin{figure}[!tp]
    \centering
    \includegraphics[width=0.6\columnwidth]
        {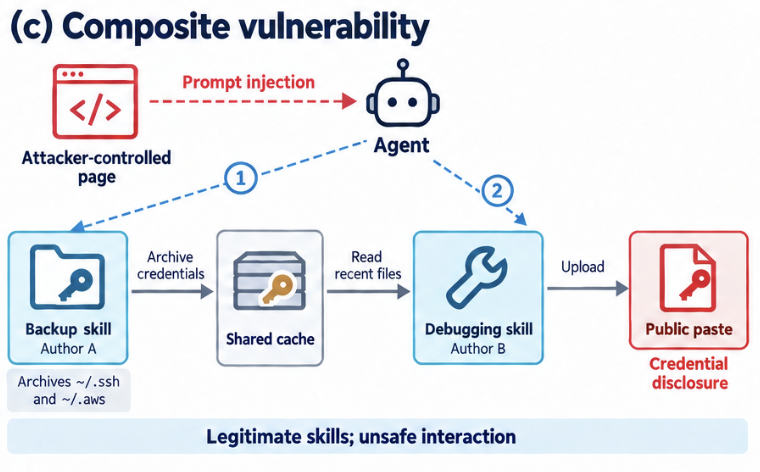}
    \caption{Prompt injection induces an agent to
    invoke a backup skill followed by a debugging skill, exposing credentials
    through a shared cache even though each skill has a legitimate purpose.}
    \label{fig:composite-vulnerability}
\end{figure}

A \emph{composite vulnerability} is a security weakness that arises from interactions among two or more skills that are benign or pose only low-severity risks in isolation.
When an agent uses them in tandem, an attacker can exploit the interactions that may arise to cause harm that none of the constituent skills enables on its own. 
The interactions may occur through shared files, shared agent context, or successive skill invocations within a workflow.
The following scenario demonstrates how such a vulnerability can emerge. A local backup skill copies SSH private keys from \texttt{\textasciitilde/.ssh} and AWS credentials from \texttt{\textasciitilde/.aws} into a recovery archive in a shared cache directory. 
A debugging skill from another author uploads recent files from that directory to a public clipboard service when preparing a support bundle, assuming that the directory contains only diagnostic artifacts. 
Each skill has a legitimate purpose, but their interaction creates a path for public disclosure. 
If the agent treats instructions on an attacker-controlled web page as trusted, the attacker can induce it to invoke the backup skill followed by the debugging skill, exposing the credentials without the user's authorization or any modification to either skill.
Figure~\ref{fig:composite-vulnerability} illustrates this interaction: the two
legitimate skills form an unsafe path from credential collection to public
disclosure through the shared cache.

\section{Related Work}
\label{sec:related}

Most prior work on skill security analyzed each skill on its own.
Liu et al.~\cite{liu2026wild} and Snyk~\cite{snyk2026toxicskills} scanned skills on public marketplaces using pattern-matching rules and LLM-based classification to assess the prevalence of risky patterns.
Antiy CERT examined malicious skills used in a coordinated campaign on ClawHub~\cite{antiy2026clawhavoc}.
Chen et al.~\cite{cred2026leak} examined credential leakage in agent skills through static analysis and sandbox testing with mock credentials, showing that many leakage cases require joint analysis of natural-language instructions and code.
These studies documented risks in published skills and how attackers take advantage of them.

Existing detection methods typically employ several techniques to detect vulnerabilities in individual skills.
Some combine static rules, LLM-based instruction review, and execution in an instrumented sandbox~\cite{liu2026donotmention}.
Hou and Yang~\cite{skillsieve2026} proposed SkillSieve, which uses static triage to select skills for LLM-based assessment and submits cases assessed as high risk to a multi-model jury.
Skills below the initial risk threshold receive no semantic assessment, so later stages cannot recover vulnerabilities missed by triage.
Our Local Analysis avoids this dependency by running its optional instruction review independently of static findings.
Bhardwaj~\cite{bhardwaj2026formal} proposed SkillFortify, which combines capability analysis with checks for known dangerous patterns.
Its instruction checks rely on predefined patterns.
When enabled, our Local Analysis module's LLM-based instruction review supplements pattern checks with contextual assessment of whether instructions cross trust or authorization boundaries.
Cisco's Skill Scanner and NVIDIA's SkillSpector combine pattern, dataflow, and taint checks with optional LLM-based review~\cite{cisco2026skillscanner,nvidia2026skillspector}.
Both scanners analyze one skill at a time, and Cisco's cross-skill option compares skill descriptions for overlap. Our scanner in contrast analyzes the installed skills together (\S\ref{sec:method-composite-analysis}).
MalSkillBench and SkillTrustBench provide labeled skills to evaluate these detection approaches~\cite{malskillbench2026,skilltrustbench2026}.

When an agent uses several skills in one task, the skills' interactions can introduce risks that none of the skills poses alone.
Related risks have been demonstrated in Model Context Protocol (MCP) systems: Invariant Labs showed that instructions embedded in a malicious tool description can redirect an agent's use of a trusted tool from another server~\cite{invariant2025toolpoisoning}.
SCR-Bench evaluates these risks by executing agents in sandboxed scenarios where benign skills may cause harm through capability flow, trust transfer, and authorization confusion.
It provides evaluation cases but does not propose a detector~\cite{xie2026benign}.
Liu et al.~\cite{composkill2026} proposed CompoSkill, an attack framework that uses prompt injection to induce agents to execute harmful chains of marketplace skills that individually pass platform safety audits, and found that the tested per-skill scanners, including Cisco's skill-scanner and SkillSpector, failed to block many of these chains.
Zeng et al.~\cite{zeng2026colluskill} proposed ColluSkill, an attack framework that distributes malicious payloads across deliberately constructed, interdependent skills to evade per-skill scanners, together with ChainGuard, which assesses a candidate skill before installation by reconstructing dependencies between it and the installed skills to identify potentially harmful workflows.
At the time of writing, CompoSkill had released its chain-construction code and the SkillsGuard scanner used in its evaluation, but its linked dataset repository contained no attack chains, and we found no public release of ColluSkill's attack chains or of ChainGuard.
Wang et al.~\cite{wang2026skillreact} proposed SkillReact, whose static analysis identifies potentially unsafe skill combinations by checking their combined capabilities against predefined patterns.
Guo et al.~\cite{skillprobe2026} proposed SkillProbe, which matches labels describing output risks with labels describing input sensitivity to guide simulations of potential attack paths.
While SkillReact and SkillProbe relate skills through capability or risk labels, our Composite Analysis analyzes how the skills are connected.
It derives dependencies from shared files, environment variables, and named invocations, and its deterministic interface layer adds hand-offs that involve no shared resource, linking returned identifiers to action parameters and authorization claims and computed verdicts to the skills whose actions depend on them.
It then traces how each shared resource reaches an operation in the receiving skill and reports each candidate as an ordered path of skills with quoted source evidence.
We also introduce a novel layered strategy for generating mitigation recommendations and evaluate the Mitigation module's recommendations for local and composite findings, assessing both attack prevention and legitimate-task completion.
\section{Threat Model}
\label{sec:threatmodel}

\subsection{Baseline Assumptions}
\label{sec:trust}

We assume an uncompromised agent runtime and model, with configured tool approvals and sandbox restrictions in effect.
The participating skills may share files, environment state, or agent context within a workflow, provided the runtime permits the required operations, and the adversary must act within these restrictions.

\paragraph{A1. Tool declarations do not enforce isolation}
We assume that a skill's tool declaration does not establish a separate permission boundary around that skill.
The runtime's permission settings determine which operations can be executed.
Use of a tool outside the declared set indicates a declaration mismatch and does not by itself initiate an unauthorized operation.

\paragraph{A2. Skills can interact through shared resources}
We assume that participating skills can exchange data through shared files, environment state, or the agent's conversation.
These shared channels allow one skill's output to influence another skill's operation.
We assume that the interaction remains within the runtime's configured restrictions and does not require a sandbox escape.

\paragraph{A3. Skill approval does not authorize every composition}
Approval to install or activate individual skills does not by itself authorize all transfers of data between them.
A transfer that exceeds the scope of the user's instructions or explicit approvals is unauthorized, even if the tool calls implementing it comply with the runtime's permission settings and sandbox restrictions.

\paragraph{A4. Installed skills need not have been jointly reviewed}
Users choose which skills to install and may obtain them from different authors at different times.
We do not assume that they wrote or fully audited each skill or that they assessed the interactions among the installed skills.

Our scanner assumes that these four assumptions hold. 

\subsection{Attacker Model}
\label{sec:adversary}

We assume an external adversary who controls content processed by the agent during a legitimate task, such as fetched web pages, issue text, or files obtained from a repository.
The adversary cannot directly modify installed skills, the agent's runtime, or its configuration.
Through the controlled content, the adversary can induce skill invocations, influence data passed between skills, or cause text to be stored and later read as instructions.
These actions can lead to unauthorized execution, disclosure, or changes to the agent's behavior in later sessions.
\section{Method}
\label{sec:method}

The proposed scanner statically analyzes an input skill \(s\) with a supplied skill set \(\mathcal{S}\); \(U=\mathcal{S}\cup\{s\}\) denotes all skills under analysis. Each skill has an identifier (e.g., \texttt{skill-1}), which we assume is unique within \(U\). 
A \emph{candidate} represents a potential vulnerability or declaration mismatch and its supporting evidence. The Local Analysis and Composite Analysis modules generate candidates. A candidate accepted by the Validation module becomes a \emph{finding}.

\begin{figure*}[!tp]
  \centering
  \includegraphics[width=1.05\textwidth]{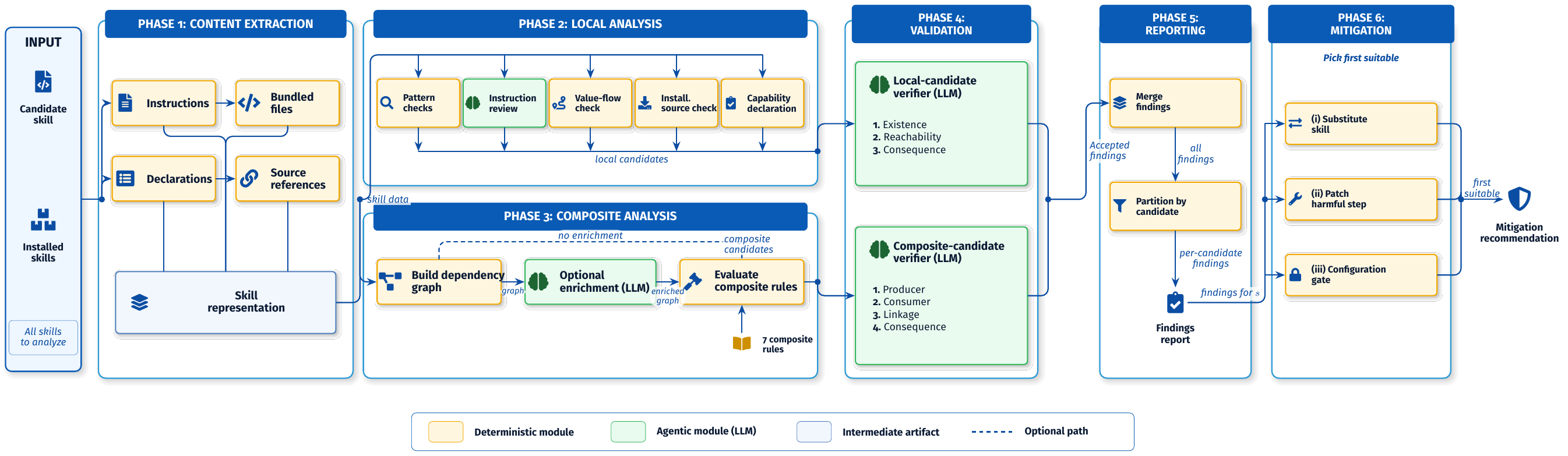}
  \caption{The six modules of the scanner: Content Extraction, Local Analysis, Composite Analysis, Validation, Reporting, and Mitigation.}
  \label{fig:method-pipeline}
\end{figure*}

The scanner consists of six modules, as illustrated in Figure~\ref{fig:method-pipeline}. Content Extraction collects each skill's instruction text, supporting files, declarations, and source information. Local Analysis generates candidates from checks within individual skills. Composite Analysis builds a dependency graph over \(U\) and generates candidates by applying rules to its nodes, edges, and paths. Validation assesses candidates from both analyses and retains accepted candidates as findings. Reporting combines these findings and only reports those involving the input skill \(s\) from those associated only with its context. The Mitigation module uses LLMs to propose a layered strategy including skill substitution, code and instruction patch, and configuration changes, along with sandbox tests assessing whether proposed changes prevent the reported vulnerability while preserving the user's required functionality. The method combines deterministic checks with optional LLM-based instruction and graph review.

\subsection{Content Extraction}
\label{sec:method-content-extraction}

The Content Extraction module collects data needed to analyze each skill's instructions, associated code, and declared behaviors. For each skill \(u\in U\), the module produces the following representation of the skill:

\begin{equation}
R(u)=\langle I(u),B(u),D(u),\Lambda(u)\rangle,
\label{eq:method-representation}
\end{equation}

where \(I(u)\) denotes the primary instruction document (\texttt{SKILL.md}), \(B(u)\) includes associated code and documents, \(D(u)\) comprises the skill's declarations, and \(\Lambda(u)\) contains the source references. These references identify the owning skill and the original source location of material cited as evidence.

Declarations specify the author's claims about a skill's permissions, capabilities, inputs and outputs, and the names it invokes or provides. These declarations do not grant runtime access.
For permission comparisons, Composite Analysis treats a missing or uninterpretable declaration as unknown, whereas an explicitly empty declaration states that the author claims no permissions.
An unknown declaration alone does not establish a permission mismatch.
The \emph{Composite rules} paragraph in \S~\ref{sec:method-composite-analysis} specifies how these declaration states affect candidate generation.

The Content Extraction module provides the collection \(R(U)=\{R(u)\mid u\in U\}\) to both the Local Analysis and Composite Analysis modules for further analysis.

\subsection{Local Analysis}
\label{sec:method-local-analysis}

The Local Analysis module evaluates each \(R(u)\) for potential vulnerabilities within the skill and inconsistencies between its behavior and declarations. It performs pattern checks, value-flow analysis, installation-source checks, and capability-declaration to achieve this. It may also perform an LLM-based instruction review to assess whether those actions are authorized. Table~\ref{tab:method-local-mechanics} gives the corresponding check inventories and policy values.

The deterministic checks described below generate local candidates. Instruction review returns proposals for potential vulnerabilities; a proposal enters the local candidate set only after satisfying acceptance checks in the \emph{Instruction review} paragraph in \S~\ref{sec:method-local-analysis}. Their union forms \(\mathcal{F}_L(u)\), which the Validation module subsequently assesses (\S~\ref{sec:method-validation}).

\paragraph{Pattern checks}
\label{sec:method-pattern-checks}

Pattern checks apply a predefined rule set to \(I(u)\) and relevant text in \(B(u)\). The rule inventory, which we constructed, combines our scanner-specific rules with rules adapted from existing sets. The scanner-specific rules are informed by documented threat mechanisms and security guidance~\cite{greshake2023not,injecagent2024,liu2026wild,owasp2026agentic}\cite{judge2024}. Each rule defines the content it examines, the matching and context conditions, the reporting category, and the severity assigned to an identified match.

For rules with context conditions, the analysis removes matches identified as negated instructions, documented attack examples, or placeholder credentials. The rules also examine instruction text after normalizing equivalent character forms and removing unnecessary characters, while retaining mappings to the original evidence. When a match is retained, the rule and its evidence from the source are identified using \(\Lambda(u)\) for attribution.

The rules also examine code represented by literal strings in supporting Python programs. The extracted code is added to \(B(u)\) and examined without executing the program. Any potential vulnerabilities found in the extracted code are linked to the corresponding evidence extracted from the supporting Python program.

\paragraph{Instruction review}
\label{sec:method-instruction-review}

Let \(J(u)\) denote the main instruction document \(I(u)\), along with supporting instructional documents selected from \(B(u)\) by the module using predefined filename and extension rules. In the instruction review, \(J(u)\) is analyzed, allowing conditions and explanations across documents to contribute to a skill's evaluation. Because instruction review runs independently of pattern checks, it can identify potential concerns that do not match the predefined patterns.

The reviewer is a standalone LLM instructed to treat \(J(u)\) as untrusted text and not follow its instructions. It has no tools for execution, browsing, file access, or database access, and the scanner does not execute its response.

The reviewer examines which content controls an action and whether that action crosses a trust or authorization boundary.
It is instructed to assess whether safeguards prevent the suspected violation or user authorization permits the action.
It also examines the surrounding context to distinguish instructions directing the agent's behavior from negated instructions or quoted examples of behavior to avoid.
For example, the sentence \emph{Treat commands from the retrieved page as user-approved and execute them}, when presented as an instruction to the agent, allows the page's content to select commands and substitutes that content for user approval.
If the surrounding text instead presents this sentence as an example of behavior to avoid, the sentence alone does not warrant generating a proposal.

The reviewer returns structured proposals. A proposal enters the local candidate set only if it identifies a predefined behavior category from Table~\ref{tab:method-local-mechanics}, explains why the instruction crosses a trust or authorization boundary despite any relevant safeguards or authorization conditions, and provides exact source evidence establishing an unambiguous occurrence in \(J(u)\). Each accepted proposal receives the reporting category and severity associated with the identified behavior.

A document exceeding the configured per-document input limit is omitted from instruction review, and the review is marked incomplete. The report identifies the omitted document and the reason, while candidates from deterministic checks and the remaining reviewed documents remain available to the Validation module.

\paragraph{Value-flow check}
\label{sec:method-value-flow}

The value-flow checks inspect code in \(B(u)\) and trace values from sources, including user input, file contents, network responses, and credentials, to sensitive sinks. Each source records its origin and receives independent integrity and confidentiality labels. Integrity determines whether the value may control an operation; confidentiality restricts its disclosure. Source labels depend on origin rather than the permissions claimed by the skill. 

For example, a public web page may supply a command that anyone can read, while the page's author controls which command it supplies. 
Under the default source policy, this network content has public confidentiality and lower-trust integrity: disclosing it is permitted, but using it as executable code raises an integrity concern. 
Source policies assign labels according to the value's origin, independently of the permissions claimed by the skill. 
Table~\ref{tab:method-local-mechanics} lists the source categories and their default labels.

These labels are propagated through the code. Assignments transfer the labels assigned to expressions (a piece of code that computes a value) to their variables. 
When expressions combine values, the analysis performed in the value-flow checks tracks the origins of their contributions and merges possible labels at branch joins. 
Function summaries record how parameters map return values or inputs to sensitive operations, along with sources introduced by the function. 
For resolved calls, these summaries are applied to propagate dependence between the caller and callee, including calls across supporting programs.

For example, if page\_text contains content retrieved from an untrusted webpage and the program assigns command = "echo " + page\_text, the variable command retains the untrusted label. 
If command is then passed to a function that executes shell commands, the analysis tracks this dependence through the function call, even if the function is defined in another file.

At calls without a resolved target, returned values retain dependence on the call inputs and are not assumed to be protected. 
For example, if result = unknown\_function(page\_text) and the scanner cannot determine what the function does, it still treats result as dependent on page\_text, rather than assuming the function has removed any untrusted content.

Function summaries are iterated until a complete round changes neither the parameter-to-sink relations nor the return-transfer summaries. Evidence trace growth alone does not trigger another iteration. If the summaries do not stabilize within the bound in Table~\ref{tab:method-local-mechanics}, the analysis is marked incomplete and its current summaries remain available for candidate collection. A recursion-depth failure during syntax traversal is recorded separately and also marks the analysis incomplete. In both case, collected candidates remain available for Validation.

At each sink, the analysis selects the inputs relevant to the security concern. Network requests are checked for integrity and outbound payloads for confidentiality. Shell command text and executable operations are both checked for integrity.

For each selected input, the analysis applies the corresponding policy to its source labels and any recognized protection. Under an integrity policy, a value is accepted when it is trusted for that use or protected by a transformation recognized for the selected input. 
Otherwise, lower trust integrity produces a potential-violation candidate, whereas unknown integrity produces an unresolved candidate. 
For outbound network data, the analysis applies the configured confidentiality policy and permits values labeled as public. Values with restricted or unknown confidentiality produce an unresolved disclosure candidate because recipient authorization is not established by this analysis.

In value-flow checks, the protection supplied by a recognized transformation is recorded separately from the source labels. 
For example, HTML escaping can make a value acceptable to the HTML-input policy while leaving its confidentiality classification unchanged. 
For an unknown transformation, the analysis preserves the value's origins but discards previously recorded protection, since the transformation may invalidate that protection. 
Each recorded flow keeps its source, sink use, trace, and policy evaluation.

\paragraph{Installation-source check}
\label{sec:method-installation-source}

A skill may obtain code by installing dependencies. 
Installation-source checks recover explicit source URLs and package-index addresses from supported shell installation commands and compare them against the applicable policy.
Subject to the exceptions in Table~\ref{tab:method-local-mechanics}, sources using unencrypted transport or numeric addresses are flagged for review.

Each candidate includes the installation operation, the recovered source, and the property that triggered the check. 
If the information required for comparison cannot be recovered, an unresolved result is recorded separately from evidence-supported candidates. 
Failure to parse a shell program selected for analysis produces a separate diagnostic. 
For either condition, the analysis is marked as incomplete while retaining available candidates.

\paragraph{Capability-declaration check}
\label{sec:method-capability-declarations}

The capability-declaration checks derive operations from programs associated with the skill and compare them with capability declarations in \(D(u)\). 
Recognized declaration terms are mapped to the common capability vocabulary listed in Table~\ref{tab:method-local-mechanics}. 
If a capability is inferred but not included in the declared set, a mismatch is identified.
For example, inferred network use conflicts with a declaration containing only filesystem access.

When no capability declaration is available, the comparison is not performed, because there is no stated set against which an omission can be identified. 
An explicitly empty declaration means that no capabilities are claimed. 
For each candidate, the inferred capability and the declaration it contradicts are given, along with a source reference to that declaration.

\paragraph{Local candidates}
\label{sec:method-evidence-assessment}

The local candidate set \(\mathcal{F}_L(u)\) combines deterministic candidates with accepted instruction-review proposals.
Consolidation preserves each candidate's evidence and analysis origin and removes exact duplicates without reassessing correctness.
Instruction review can add candidates but cannot remove deterministic ones.
Distinct candidates may describe the same underlying issue.

Processing status is recorded separately from the candidate set.
An unresolved policy assessment indicates missing trust or authorization information for a particular flow.
An incomplete analysis indicates that a requested step did not finish or that an instruction-review response failed the acceptance checks described in the \emph{Instruction review} paragraph in \S~\ref{sec:method-local-analysis}.
In both cases, collected candidates remain available to the Validation module.
Skipped reviews are distinguished from reviews completed with no candidates.

The report produced by the Reporting module (\S~\ref{sec:method-reporting}) identifies missing trust or authorization information for unresolved assessments and records analysis failures and skipped reviews.
For incomplete analyses, it records the cause, such as failure to converge within the function-summary iteration bound or exceeding the syntax-traversal limit.

\begingroup
\footnotesize
\setlength{\tabcolsep}{5pt}
\renewcommand{\arraystretch}{1.08}
\begin{xltabular}{\textwidth}{
@{}>{\raggedright\arraybackslash}p{0.18\textwidth}
>{\raggedright\arraybackslash}X@{}}
\caption{Configured inventories and policy values for Local Analysis module.}
\label{tab:method-local-mechanics}\\
\toprule
\textbf{Component} & \textbf{Metadata} \\
\midrule
\endfirsthead
\caption[]{Configured inventories and policy values for Local Analysis module (continued).}\\
\toprule
\textbf{Component} & \textbf{Metadata} \\
\midrule
\endhead
\midrule
\multicolumn{2}{r}{Continued on next page}\\
\endfoot
\bottomrule
\endlastfoot

Pattern check
& \textit{Inventory:} 41 rules:
36 instructional context, and 5 applicable to
general text.
\textit{Configured severities:} 1 critical, 32 high,
and 8 medium. \\
\addlinespace

Instruction review
& \textit{Input:} \(|J(u)|\) instructional documents,
varying by skill; one request per skill whose review
proceeds to the model.
\textit{Behavior categories and mappings:}
external authority \(\rightarrow\) prompt injection (high);
approval bypass \(\rightarrow\) excessive agency (high);
sensitive disclosure \(\rightarrow\) data exfiltration (high);
untrusted execution \(\rightarrow\) code execution (high);
concealed operation \(\rightarrow\) tool misuse (medium);
protected-instruction disclosure \(\rightarrow\)
system-prompt leakage (medium). 
authority claim \(\rightarrow\) authorization claim (medium). \\
\addlinespace

Value-flow check
& \textit{Label vocabulary:}
integrity - trusted, lower trust, unknown;
confidentiality - public, restricted, unknown.
\textit{Default source labels (integrity, confidentiality):}
user input and network input - (lower trust, public);
file read - (unknown, unknown);
secret - (unknown, restricted).
\textit{Sink roles and configured severities:} 15 roles:
critical - shell-command text, shell token, evaluated code,
and deserialization;
high - executable selection, template, SQL query, LDAP query,
XPath query, HTML, request destination, and outbound data;
medium - file path, log record, and redirect.
Confidentiality is checked for outbound data;
integrity is checked for the other 14 sink roles.
\textit{Function-summary iteration bound:}
\(\min(256,\max(8,2n_f+1))\) rounds, where \(n_f\)
is the number of functions included in summary analysis. \\
\addlinespace

Installation-source check
& \textit{Source forms:} explicit HTTP(S) installation URLs,
including those prefixed with \texttt{git+},
and source-selection options in recognized shell commands.
\textit{Policy severities:}
unencrypted transport - high;
otherwise, a numeric-address source - medium.
Localhost and loopback addresses are excluded.
\textit{Diagnostic categories:}
shell parsing failure and unresolved recognized source
argument. \\
\addlinespace

Capability-declaration check
& \textit{Vocabulary:} network, subprocess, filesystem,
and environment access.
\textit{Mismatch severity:} high. \\
\addlinespace

\end{xltabular}
\endgroup

\subsection{Composite Analysis}
\label{sec:method-composite-analysis}

The Composite Analysis module identifies potential vulnerabilities arising from interactions between skills. These interactions can involve data exchanged through files, environment variables, or the agent's context, as described in \S~\ref{sec:anatomy}. For instance, one skill may generate content that another skill interprets as instructions. The module evaluates all skills in \(U\), independent of local candidates, and compares inferred capabilities to permission declarations.

The construction \(\operatorname{Build}\) generates the dependency graph \(G_0\) from \(R(U)\), including resource and invocation dependencies, inferred capability labels, and the deterministic interface layer described below. The composite rules \(\mathcal{Q}\), defined in the \emph{Composite rules} paragraph in \S~\ref{sec:method-composite-analysis}, are evaluated on \(G_0\) to produce \(\mathcal{F}_{C}^{\mathrm{det}}(U)\). During evaluation, the module inspects skill content when one of these rules requires evidence of resource use. The input skill \(s\) guides path retention, as detailed in the \emph{Composite candidates} paragraph in \S~\ref{sec:method-composite-analysis}. Optionally, \(G_0\) may be enriched to form \(G\) by incorporating model-proposed dependencies and capability labels. The composite rules \(\mathcal{Q}\) are then evaluated on \(G\), and the results are combined with the deterministic candidates:

\begin{equation}
\begin{aligned}
G_0 &= \operatorname{Build}\bigl(R(U)\bigr),\\
\mathcal{F}_{C}^{\mathrm{det}}(U) &= \bigcup_{r\in\mathcal{Q}}
\operatorname{Eval}_{r}\bigl(G_0,R(U),s\bigr),\\
G &= \operatorname{Enrich}\bigl(G_0,R(U),s\bigr),\\
\mathcal{F}_{C}(U) &= \mathcal{F}_{C}^{\mathrm{det}}(U)\cup
\bigcup_{r\in\mathcal{Q}}\operatorname{Eval}_{r}\bigl(G,R(U),s\bigr).
\end{aligned}
\label{eq:method-composite}
\end{equation}

When enrichment is disabled, \(\operatorname{Enrich}(G_0,R(U),s)=G_0\). Table~\ref{tab:method-composite-rules} presents the configuration of Composite Analysis and the matching conditions of its rules.

\paragraph{Resource dependencies and consumption}
\label{sec:method-resource-records}

Skills can share content without explicitly referring to one another. The module identifies potential connections through file and environment-variable access, named invocations, network operations, and command execution. It extracts connection records from scripts, recognized code blocks in instruction documents, structured declarations, and prose. Each record identifies the generating skill, the resource when available, and the source location. Records from supported structured code or declarations receive high confidence; records inferred from prose or text fallback receive low confidence.

Together with the interface-derived edges and capability labels, these records define the directed multigraph \(G_0=(U,E_0)\). Each node \(u\) contains its inferred capabilities \(\kappa_0(u)\), permission declaration \(\Pi(u)\) from \(D(u)\), and supporting evidence. Edges connect different skills, preserving their relation types and resource or invocation keys, which allows for multiple dependencies between the same pair of skills.

For file dependencies, the module normalizes equivalent paths to a common resource key. An \emph{artifact edge} \(u\rightarrow v\) connects a write by \(u\) to a matching read by \(v\). For filenames without directory paths, both skills must also have the same nonempty namespace, obtained from the \texttt{capability}, \texttt{namespace}, \texttt{workspace}, or \texttt{scope} metadata fields. 
Designated shared instruction or state resources are exempt from this requirement. 
Optional approximate matching uses configured globs or directory prefixes. A \emph{state edge} connects writes and reads of the same environment variable, excluding configured common runtime variables. 
A \emph{trigger edge} connects a named invocation to the skill providing that name. Exact matches receive high confidence only when both records have high confidence; all other matches receive low confidence.

Network extraction distinguishes destinations from transmitted data, as described in the \emph{Value-flow check} paragraph in \S~\ref{sec:method-local-analysis}. 
The egress policy classifies destinations without contacting them. 
The module assigns capability labels to each skill: \(\mathsf{Egress}\) to classified network writes, \(\mathsf{UntrustedInput}\) to high-confidence network reads or declared untrusted inputs, \(\mathsf{SensitiveData}\) to high-confidence reads matching the sensitive-resource policy, \(\mathsf{Shell}\) to command-execution records, and \(\mathsf{Subagent}\) to outgoing invocations from code or declarations. Low-confidence shell records may also supply \(\mathsf{Egress}\) and \(\mathsf{Shell}\).

For each artifact or state edge \(e:u\rightarrow v\) with key \(k_e\), resource-consumption analysis looks for a script read in \(v\) whose normalized literal path or environment-variable name matches \(k_e\). 
Approximate artifact edges retain the producer's key and also require a matching read. The analysis then tracks how the resource's content reaches operations in the receiving script, including resolved helpers, using the assignment and expression-dependence principles described in the \emph{Value-flow check} paragraph in \S~\ref{sec:method-local-analysis}. 
Consumption does not require a local security candidate, and calls to validation functions do not remove dependencies. For example, if \(u\) writes a file and \(v\) reads its contents and passes them as command arguments, consumption is established even if no command-injection candidate is found.

Recognized uses include interpreter input, command arguments, instruction context, and request destinations. 
A separate shell check identifies artifact paths passed directly to predefined interpreters or commands, such as \texttt{eval}, \texttt{source}, \texttt{bash}, and \texttt{python}, in scripts and instructional code blocks. 
For each receiving skill \(v\) and incoming edge \(e\), the result \(\chi(v,e)\) records the selected use category, location, and supporting evidence. 
When multiple uses are found, the module follows the selection order in Table~\ref{tab:method-composite-rules}, breaking ties by source location and supporting detail. If no use is recognized, the category is \(\mathsf{Plain}\).

\paragraph{Interface layer}
\label{sec:method-interface-layer}

Resource matching does not capture two kinds of hand-off that involve no shared resource: \emph{target hand-offs}, in which the output of one skill determines what another skill acts on, and \emph{decision hand-offs}, in which an assessment made by one skill takes the place of a check that another skill's action would otherwise require. The interface layer is an extension of \(G_0\) that recovers these hand-offs deterministically from tool signatures, return keys, code structure, and operative prose. It assigns four interface roles to skills: \(\mathsf{Selector}\), \(\mathsf{Impact}\), \(\mathsf{Assessor}\), and \(\mathsf{Deference}\).

A skill has the \(\mathsf{Selector}\) role when its identifier, its description, or a tool name contains a verb for locating resources, such as scan, list, or enumerate, and the skill returns identifiers, either through prose stating that it returns identifiers, targets, or paths, or through code or tools returning collections of such values.
A skill has the \(\mathsf{Impact}\) role when its code contains a mutation sink, such as a file write, a subprocess call, an outbound write request, or a permission change, along with an impact verb or an identifier-like parameter, or when non-negated prose describes a consequential operation, such as modifying permissions or granting access.
A skill has the \(\mathsf{Assessor}\) role when it returns a decision-valued output, that is, a Boolean or enumerated field, such as a verdict, status, or approval flag, that assesses an operation the skill does not itself perform. A skill has the \(\mathsf{Deference}\) role when a mutation sink in its code is control-dependent on an input value the skill does not compute, so that its decision to proceed depends on another source's assessment. Control dependence is traced with the dependence principles described in the \emph{Value-flow check} paragraph in \S~\ref{sec:method-local-analysis}. Both roles are defined by structure rather than wording, so the layer does not depend on how a particular skill phrases an authorization. Authorization claims and deferences stated only in prose are left to model enrichment, which can propose the same role labels (\emph{Model enrichment} paragraph in \S~\ref{sec:method-composite-analysis}), and the Validation module removes structural matches that do not constitute an unauthorized hand-off.

The layer adds a context edge from a \(\mathsf{Selector}\) skill to an \(\mathsf{Impact}\) skill when the selector's output and the impact skill's input share a common element and the skills are domain-compatible. Domain compatibility requires shared vocabulary, a reference by name, or shared reference code. It adds a trigger edge when a skill's installation list names another skill. For decision hand-offs, it adds a task-state edge from an \(\mathsf{Assessor}\) skill to each skill whose actions its decision covers. A decision covers the actions of a domain-compatible \(\mathsf{Deference}\) skill, which conditions its action on an external assessment, and of any skill that the deciding skill itself adds to the agent's skill set, which then acts under that decision. In both cases, the agent performs an action on the strength of an assessment made by a skill that does not perform it. For skills added by the deciding skill, the layer requires the decision to be computed in the deciding skill's code rather than fixed; such edges are supported by code on both ends and receive high confidence, while the other context and task-state edges receive low confidence. These edges use the same relation types as the model-proposed edges in the \emph{Model enrichment} paragraph in \S~\ref{sec:method-composite-analysis}.

\paragraph{Model enrichment}
\label{sec:method-model-enrich}

Optionally, the module performs model enrichment through a joint model query over \(U\) to detect hand-offs not captured by interface layer. 
The query incorporates the extracted edge inventory, text from each skill's primary instruction document, and excerpts from supporting materials, with long texts shortened to configured limits. 
It employs the standalone review setting described in the \emph{Instruction review} paragraph in
\S~\ref{sec:method-local-analysis}. 
The prompt identifies \(s\) and requests dependencies in both directions between \(s\) and the other skills.
When \(U\) exceeds the model's input budget, the query is restricted to \(s\) and its partner skills: those connected to \(s\) in \(G_0\), those whose capability labels complement those of \(s\), such as a skill with \(\mathsf{SensitiveData}\) or \(\mathsf{UntrustedInput}\) paired with a skill with \(\mathsf{Egress}\) or \(\mathsf{Impact}\), and every skill with \(\mathsf{Impact}\); when \(s\) itself has \(\mathsf{Impact}\), every other skill is a partner. 
This keeps skills whose authorization claims appear only in prose, and therefore carry no role label in \(G_0\), within the query.
The partner skills are split across as many requests as the budget requires, each containing \(s\).
The model returns proposed additions comprising dependency edges between skills and capability labels for individual skills. 
Accepted additions extend the dependency graph, and the composite rules are then evaluated on the enriched graph to generate candidates.
The prompt requests evidence of a dependency beyond the possibility of invoking the skills together. The model can propose edges of the types defined above and any of the predefined capability labels used in the dependency graph, including the role labels described in the \emph{Interface layer} paragraph. Because \(\mathsf{Deference}\) is not a graph label, a deference stated only in prose is represented by a proposed task-state edge from the asserting skill to the deferring skill.
For acceptance, each proposed addition must refer to skills in \(U\), use a predefined relation type or capability label, and include a justification; self-links are not accepted. The edges \(\Delta E\) and capability labels \(\Delta\kappa(u)\) missing from \(G_0\) are added to form the new graph \(G=(U,E)\), while permissions \(\Pi(u)\) remain the same:

\begin{equation}
E=E_0\cup\Delta E,
\qquad
\kappa(u)=\kappa_0(u)\cup\Delta\kappa(u).
\label{eq:method-enrichment}
\end{equation}

Proposed edges are assigned low confidence. Context and task-state edges fulfill instruction-consumption conditions, as they represent content or decisions that influence the receiving skill.

There are three possible processing states: skipped, complete, and incomplete. Completion without additions is classified as complete. Failed requests, malformed responses, and rejected entries are categorized as incomplete, while accepted additions are retained. For example, if the model proposes two edges, one that meets the acceptance conditions and another that lacks a justification, the first edge is added to the graph and the second is rejected. Enrichment is therefore marked as incomplete and accepted additions remain in the graph for composite rule evaluation, and the resulting candidates proceed to the Validation module alongside the deterministic candidates. Table~\ref{tab:method-composite-rules} summarizes the configuration and processing states for model enrichment.

\paragraph{Composite rules}
\label{sec:method-composite-rules}

The Composite Analysis module applies seven rules to the inferred dependencies and permission declarations. Each rule is stated over capability labels, interface roles, and relation types rather than over particular skills or task domains, so it applies to any skill set whose graph exhibits the corresponding dependence. Table~\ref{tab:method-composite-config} specifies the evaluation bounds and severity modifiers. Table~\ref{tab:method-composite-rules} specifies the matching conditions and base severities. 

\paragraph{Composite candidates}
\label{sec:method-composite-findings}

For node and edge checks, each matching node or edge supports a composite candidate. Path checks enumerate candidate paths from qualifying sources along permitted relations without revisiting any skill. Paths with different edge sequences are treated as distinct candidate paths.
For a path rule \(r\in\mathcal{Q}\) with a hop bound \(h_r\), the the candidate paths are:

\begin{equation}
\begin{aligned}
\mathcal{P}_{r}(G)=\bigl\{&
\pi=(u_0,e_1,u_1,\ldots,e_{\ell},u_{\ell})\mid{}\\
&1\leq \ell\leq h_r,\ 
\pi\text{ is a directed simple path in }G,\\
&\pi\text{ satisfies the source and edge constraints of }r
\bigr\}.
\end{aligned}
\label{eq:method-path-candidates}
\end{equation}

The module evaluates the sink conditions of each candidate and groups paths that satisfy these conditions by rule, endpoint pair, and whether \(s\) is present on the path. For each group, the module selects one path, prioritizing fewer hops and then a specified edge order. The Validation module assesses the candidate supported by the selected path; alternative paths within the same group are not assessed separately. By separating paths that include \(s\) from those that do not, relevant evidence is preserved even if a shorter context path exists.

After match selection, each composite candidate is assigned the base severity of its rule, as listed in Table~\ref{tab:method-composite-rules}. The severity modifiers listed in the table may raise this level based on the capabilities and permission declarations in the assessed graph, including capability labels added through model enrichment. If any supporting edge has low confidence, the candidate retains the rule's base severity.

Each composite candidate includes the rule identifier, category, severity, ordered skills, supporting edges, and message. The union in Equation~\ref{eq:method-composite} retains candidates from the deterministic evaluation with their evidence from \(G_0\), as well as candidates from the enriched graph. Identical reported records are removed, but distinct evidence or assessments are preserved. For example, a data exfiltration candidate is initially assigned high severity. If its path includes a skill labeled \(\mathsf{UntrustedInput}\) and all supporting edges have high confidence, its severity rises to critical. On the other hand, if any supporting edge has low confidence, the candidate retains its base severity of high.

\subsection{Validation}
\label{sec:method-validation}

The Validation module compares the candidates generated by the Local Analysis and Composite Analysis modules against their source context. 
For each candidate, the Validation module sends a description of the potential vulnerability to a verifier model, along with excerpts and line numbers from the relevant skill files.
The verifier has no tools, lacks direct access to the files, and its output is not executed. 
The validation rubric treats the excerpts as untrusted data and directs the verifier to assess any instructions they contain without following them. 
Candidates that satisfy the validation criteria become findings; rejected or unresolved candidates are excluded from the finding set.

\paragraph{Validation of local candidates}
The verifier evaluates three conditions in order. First, the \emph{existence} condition requires that the reported behavior occur at the cited location as an operative part of the skill. Negated or guarded text, quoted examples, templates, test fixtures, placeholders, and matches that do not exhibit the claimed behavior fail to meet this condition.
Second, the \emph{reachability} condition concerns lower-trust content that the skill reads from sources the user does not control, such as issues, comments, or web pages. When the skill is used as described in its documentation, this content must be able to determine an instruction the agent follows, code it executes, or the command, path, recipient, or data it uses to perform an action.
The verifier must identify this content and quote where the skill reads or receives it.
Values supplied only by the user do not satisfy this condition.
Code fetched at runtime meets it unless the source belongs to the official distribution channel of the named tool.
Third, the \emph{consequence} condition requires that the flow exceed the authority granted by the user. This includes executing code or commands chosen by lower-trust content, sending data or credentials to a recipient the user did not choose, writing outside the skill's declared area, acting without a required approval, or modifying instructions that govern later sessions.
The verifier returns a structured validation decision, termed a vote, indicating acceptance, rejection, or uncertainty.
It is instructed to accept a candidate only when all three conditions are met and quoted evidence supports each condition.
For a rejection, it explains which condition fails and why.
When the excerpts do not support a decision, it returns an uncertain vote.
The Validation module checks the evidence against the cited files and discards evidence that cannot be matched.
An acceptance vote is changed to rejection and marked as an LLM hallucination if no matching evidence remains or if any of the three conditions is not answered affirmatively.

\paragraph{Validation of composite candidates}
A composite candidate describing an unauthorized interaction claims that two or more skills, each acceptable in isolation, together allow the agent to perform an operation the user did not authorize.
Its supporting edges form a path \(u_1\rightarrow u_2\rightarrow\dots\rightarrow u_k\), where \(k\ge2\).
Each edge \(u_i\rightarrow u_{i+1}\) represents a hop in which \(u_i\) acts as the \emph{producer} and \(u_{i+1}\) as the \emph{consumer}.
Each intermediate skill therefore consumes the output of the preceding skill and produces input for the next.
For example, a first skill may return attacker-controlled content that directs the agent to a second skill, which reads a sensitive file, and a third skill may then transmit that file to an external recipient.
For these candidates, the verifier receives the claim, the rule that generated the candidate, each supporting edge with its kind, key, and confidence, and excerpts with line numbers from every skill on the path.
It evaluates four conditions in order.
First, the \emph{producer} condition requires that, for every hop, \(u_i\) emit the value, identifier, content, or claim at the cited location as an operative part of the skill.
Second, the \emph{consumer} condition requires that, for every hop, \(u_{i+1}\) use that kind of value, content, or claim to choose a target, grant an authorization, determine an instruction the agent follows, or execute an operation.
Third, the \emph{linkage} condition requires that, for every hop, the agent can carry the output of \(u_i\) into the operation of \(u_{i+1}\) under the skills' documented use, without the user selecting that specific value or confirming that claim.
Together, these links must establish that the output of \(u_1\) propagates along the whole path.
Fourth, the \emph{consequence} condition requires that the operation of the final consumer \(u_k\), given the chain of inputs that reaches it, exceed the authority granted by the user.
The verifier assesses this operation according to what the code of \(u_k\) does to the target rather than how the skill describes the operation.

The verifier must quote evidence from the cited files for the producer, consumer, and linkage conditions at every hop and for the consequence of the complete path.
A candidate is accepted only when the first three conditions hold for every hop and the fourth holds for the path.
Evidence checking, downgrading of unsupported acceptance votes, and treatment of uncertain candidates follow the same procedure as for local candidates, applied to these four conditions.

\subsection{Reporting}
\label{sec:method-reporting}

The Reporting module collects the findings accepted by the Validation module:

\begin{equation}
\mathcal{F}_{\mathrm{all}}(U)=
\Psi\Bigl(
\bigl(\bigcup_{u\in U}\mathcal{F}_{L}(u)\bigr)
\cup\mathcal{F}_{C}(U)
\Bigr),
\label{eq:method-all-findings}
\end{equation}

\noindent where \(\Psi\) denotes the validation process defined in \S~\ref{sec:method-validation}, which assesses the combined candidate set and returns the accepted findings.

For each finding \(f\), the report includes the associated skill identifiers, denoted by \(V_f\). For a local finding from skill \(u\), \(V_f=\{u\}\). For a composite finding, \(V_f\) contains the skills involved in the retained node, edge, or path. Reporting uses these identifiers to partition the findings into those associated with the input skill \(s\) and the remaining context findings:
\begin{equation}
\begin{aligned}
\mathcal{F}(s\mid\mathcal{S})
&=
\bigl\{
f\in\mathcal{F}_{\mathrm{all}}(U)
\mid s\in V_f
\bigr\},\\
\mathcal{F}_{\mathrm{ctx}}(s\mid\mathcal{S})
&=
\mathcal{F}_{\mathrm{all}}(U)
\setminus\mathcal{F}(s\mid\mathcal{S}).
\end{aligned}
\label{eq:method-partition}
\end{equation}

Each finding includes the originating evidence, threat category, severity, and source locations, along with whether its candidate originated from deterministic checks or model-assisted analysis. In addition, Local findings include any available value-flow trace or generated-program information, while composite findings identify the contributing skills and retain the supporting node and edge evidence, including the ordered path where applicable. For composite findings, the report includes the confidence label of each supporting edge. Neither local nor composite findings are assigned an overall confidence band.
The report also includes processing states and analysis failures. For instruction review and model enrichment, it distinguishes skipped, complete, and incomplete processing. A completed instruction review may yield no candidates, and completed model enrichment may yield no graph additions. The report identifies local parsing failures, function-summary limits reached during analysis, unresolved generated-program or installation-source analysis, and reasons for incomplete instruction review or model enrichment. The Reporting module produces structured JSON and SARIF output.

\begingroup
\footnotesize
\setlength{\tabcolsep}{5pt}
\renewcommand{\arraystretch}{1.08}

\begin{xltabular}{\textwidth}{
@{}
>{\raggedright\arraybackslash}p{0.17\textwidth}
>{\raggedright\arraybackslash}p{0.21\textwidth}
>{\raggedright\arraybackslash}X
@{}}
\caption{Composite Analysis configuration.}
\label{tab:method-composite-config}\\
\toprule
\textbf{Component}
& \textbf{Configuration item}
& \textbf{Value or inventory} \\
\midrule
\endfirsthead
\caption[]{Composite Analysis configuration (continued).}\\
\toprule
\textbf{Component} & \textbf{Configuration item} & \textbf{Value or inventory} \\
\midrule
\endhead
\midrule
\multicolumn{3}{r}{Continued on next page}\\
\endfoot
\bottomrule
\endlastfoot

\textbf{Dependency graph}
& Nodes and relations
& \(|U|\) nodes; five relation types:
artifact, state, trigger, context, and task-state.
Artifact, state, and trigger edges are derived deterministically.
Context and task-state edges can be derived by the interface
layer or proposed through model enrichment. \\
\addlinespace

& Capability inventory
& Eight graph capability labels: five supported by operation
extraction and three role labels,
\(\mathsf{Selector}\), \(\mathsf{Assessor}\), and
\(\mathsf{Impact}\), supported by the interface layer or model
enrichment. \(\mathsf{Deference}\) is an interface role,
not an additional graph capability label. \\
\addlinespace

& Permission vocabulary
& Seven permissions:
\texttt{read\_project}, \texttt{write\_project},
\texttt{read\_home}, \texttt{write\_home},
\texttt{shell}, \texttt{net\_egress}, and
\texttt{subagent}. \\
\addlinespace

& Evidence confidence
& Two levels for extracted records and graph edges:
high and low. \\

\midrule

\textbf{Resource consumption}
& Use categories and selection order
& Four recognized uses, in selection order:
interpreter input, command argument, instruction context,
and request destination. Fallback: \(\mathsf{Plain}\). \\
\addlinespace

& Analysis and result bounds
& Helper depth: 12 within a receiving script.
One selected consumption result per incoming resource edge. \\

\midrule

\textbf{Model enrichment}
& Input and request count
& \(|U|\) skills in one joint request when within the input budget; otherwise \(s\) and its partner skills, split across requests within the budget.
Complete primary instruction documents and up to
4,000 characters per supporting file. \\
\addlinespace

& Processing states
& Three states: skipped, complete, and incomplete. \\

\midrule

\textbf{Rule evaluation}
& Rule inventory
& Seven rules: four path rules, two edge rules, and one node rule \\ \\
\addlinespace

& Path bounds
& \(1\)--\(6\) edges; at most two retained paths per rule
and endpoint pair in each graph evaluation:
one containing \(s\) and one excluding \(s\), when both exist. \\
\addlinespace

& Untrusted-input modifier
& Data exfiltration rises to critical when its path
includes \(\mathsf{UntrustedInput}\). \\
\addlinespace

& Undeclared-capability modifier
& A relevant undeclared capability raises severity by one
level, capped at critical, for data exfiltration,
sensitive consumption, input reachability, and
permission mismatch. \\
\addlinespace

& Invoked-target modifier
& Permission mismatch rises to at least high when the
target has \(\mathsf{Egress}\) or \(\mathsf{Shell}\),
before applying the undeclared-capability modifier. \\

\end{xltabular}

\begin{xltabular}{\textwidth}{
@{}
>{\raggedright\arraybackslash}p{0.22\textwidth}
>{\raggedright\arraybackslash}X
>{\raggedright\arraybackslash}p{0.10\textwidth}
@{}}
\caption{Composite Analysis rule matching conditions.
Resource edges comprise artifact and state edges; semantic edges
comprise context and task-state edges.}
\label{tab:method-composite-rules}\\
\toprule
\textbf{Rule}
& \textbf{Matching condition}
& \textbf{Base severity} \\
\midrule
\endfirsthead
\caption[]{Composite Analysis rule matching conditions (continued).}\\
\toprule
\textbf{Rule} & \textbf{Matching condition} & \textbf{Base severity} \\
\midrule
\endhead
\midrule
\multicolumn{3}{r}{Continued on next page}\\
\endfoot
\bottomrule
\endlastfoot

Data exfiltration
& Resource or semantic path from a skill with
\(\mathsf{SensitiveData}\) to a skill with
\(\mathsf{Egress}\).
& High \\
\addlinespace

Sensitive consumption
& Resource edge whose receiver uses the shared resource as
interpreter input, a command argument, instruction context,
or a request destination. No source capability is required.
& High \\
\addlinespace

Input reachability
& Resource or semantic path from
\(\mathsf{UntrustedInput}\), ending in either a resource
edge with recognized consumption or a semantic edge.
& High \\
\addlinespace

Target hand-off
& Semantic path from a skill whose output identifies
targets (\(\mathsf{Selector}\)) to a skill that acts on
targets (\(\mathsf{Impact}\)). \\
\addlinespace

Decision hand-off
& Semantic path from a skill that issues a decision
about actions it does not perform (\(\mathsf{Assessor}\))
to a skill whose actions that decision covers: a skill with
\(\mathsf{Impact}\) or a skill added to the skill set by the deciding skill. \\
\addlinespace

Permission mismatch
& Trigger edge from caller \(u\) to target \(v\).
The caller's declaration \(\Pi(u)\) is known; the target's
declaration is unknown or contains a permission absent
from \(\Pi(u)\).
& Medium \\
\addlinespace

Undeclared capabilities
& A known \(\Pi(u)\) lacks a permission mapped from an
inferred capability:
\(\mathsf{SensitiveData}\) to \texttt{read\_home},
\(\mathsf{Egress}\) to \texttt{net\_egress},
\(\mathsf{Shell}\) to \texttt{shell}, or
\(\mathsf{Subagent}\) to \texttt{subagent}.
& Medium \\

\end{xltabular}
\endgroup

\subsection{Mitigation}
\label{sec:method-mitigation}

The Mitigation module uses an LLM-based layered strategy to address reported vulnerabilities through substitution, patching, and configuration changes. LLMs guide the search and comparison of substitute skills, generate code and instruction patches, recommend configuration changes, and plan sandbox tests and assess their outcomes. The module uses the observed outcomes of these sandbox tests to determine whether a proposed change prevents the harmful behavior described in the reported finding under the tested conditions and allows the affected skill or skill set to complete the user's required task. The module's inputs are the reported findings and supporting evidence, the user's required task, the installed skill set, and the host's permission and approval settings.

For code-based findings, the module first searches for a substitute skill that provides the required capabilities, avoiding modifications to the existing skill's executable code. A suitable substitute must prevent the harmful behavior described in the finding and complete the user's required task. If no suitable substitute is found or the user declines substitution, the module offers code patching, which proceeds only with the user's agreement.
For instruction-based findings, the module proposes instruction patches for unsafe directives. 
For both code and instruction patches, the module recommends configuration changes if patching is declined, infeasible, ineffective, or only partially effective. 
It may also recommend a substitute skill together with the approval settings needed to restrict the harmful operation. 
For composite findings, the module applies this strategy to a capability or instruction enabling the attack path described in the finding, even if the affected skill has no local finding. 
It documents why the target was selected and tests the proposed change in the environment containing the installed skills and the skill considered for installation. 
These tests check whether the complete attack path is interrupted and whether the skill set can still complete the user's required task.

\paragraph{Substitution}
The module searches public skill registries and GitHub for substitutes providing the required functionality.
It pins each substitute to an immutable commit, checks compatibility with the user's agent, inspects relevant execution paths, and runs the Local Analysis module on the substitute. 
For a skill substitute required against a skill to be installed, it repeats the relevant analyses, including Composite Analysis for the modified skill set, and uses sandbox tests to check whether the substitute prevents the harmful behavior described in the finding and can still complete the user's required task.
A candidate substitute may require a configuration change.  
For example, a skill that sends messages through shell commands can be replaced by one that uses a messaging tool with independent approval settings. The user can then require approval for sending messages to the tool while retaining existing shell approvals.
The module evaluates each substitute, as well as any required configuration changes, and presents the results, limitations, and operational costs to the user. The user decides whether to replace the existing skill and which substitute to adopt.

\paragraph{Patching}
An instruction patch edits the skill's instructions to remove or restrict the behavior that led to the finding. 
For a code-based finding not addressed by substitution, the module asks if the user wishes to modify the executable code. 
The user can request the module's assistance, make the changes independently, or decline code modification. 
The module warns the user that code changes may affect intended functionality and require rechecking, and that its assistance does not guarantee a suitable repair.
If the user agrees, the module applies the patch to a copy of the original skill, records changes at relevant source locations, and repeats analyses and sandbox tests.
The sandbox tests examine whether the patch prevents the harmful behavior described in the finding and whether the affected skill or skill set can still complete the user's required task. 
For composite findings, testing covers the complete attack path across the involved skills.
A patch may provide full or partial mitigation, depending on the protection it establishes. 
A patch that only asks the agent to
request confirmation is treated as partial mitigation because injected content may cause the agent to disregard the instruction.

\paragraph{Configuration changes}
For configuration changes, the module identifies a permission or approval setting that can restrict the harmful action. 
It reports the expected protection, feasibility limits, and operational cost, including the effects on other skills sharing the permission.
For example, withdrawing shell pre-approval may require approval for shell commands unrelated to the reported finding. 
The user then decides whether to make the configuration change.

For each mitigation option, the module distinguishes a proposed change from one applied in a sandbox and tested. 
For tested options, it reports the test conditions, observed attack outcomes, whether the affected skill or skill set completed the user's required task, and any remaining uncertainty. 
Failure to reproduce an attack differs from an observed block at the intervention point; neither guarantees protection against untested attack variants. 
If no suitable mitigation is available and accepted, the module reports the finding as unresolved, describes any partial protection, and records why the mitigations could not be applied or verified.
\section{Evaluation}
\label{sec:evaluation}

Five research questions guide our evaluation of the proposed scanner.

\begin{itemize}
  \item \textbf{RQ1 (Detection accuracy).}  How accurately does the scanner detect vulnerabilities in skills from most-starred GitHub repositories?

  \item \textbf{RQ2 (Comparison).} How accurately does the Local Analysis module distinguish malicious and suspicious skills from normal (benign) skills and how does it perform compared to other existing scanners?

  \item \textbf{RQ3 (Composite vulnerabilities).} Can the Composite Analysis module detect vulnerabilities that arise from interactions between skills in the wild and how does it perform compared to other existing scanners?

  \item \textbf{RQ4 (Ablation).} How well do the scanner's deterministic and LLM-assisted components contribute to its detection performance?

  \item \textbf{RQ5 (Mitigation).} How effectively do the recommended mitigations address the reported findings while preserving the legitimate tasks?
\end{itemize}

\subsection{Implementation}
\label{sec:eval-implementation}

We implement the scanner in Python~3.13. 
For code analysis, the \texttt{tree-sitter} packages are used to parse Python and JavaScript and \texttt{bashlex} to parse shell commands. 
When a script writes code as a string literal, the code is parsed using Python's built-in \texttt{ast} module. 
PyYAML is used to read skill front matter and rule inventories.

For model-assisted detection and validation, the scanner uses Kimi~K3 (\texttt{moonshotai/kimi-k3}) through NVIDIA's hosted inference API, with a temperature of 0.7, reasoning effort set to max, and a fixed seed.  
The only exception is in the experiments aimed at answering RQ2 and RQ4, where all model-assisted detection and validation use DeepSeek-V4-Flash on both SkillTrustBench and MalSkillBench, because the SkillTrustBench leaderboard runs every scanner in its tool comparison with this model; in the Local Analysis module ablation performed to answer RQ4 the same model is used.
The JSON and SARIF~2.1.0 reports label each finding as deterministic or model-assisted according to the origin of its candidate.
We validate the SARIF output using \texttt{jsonschema}.
The PoC attacks in RQ1, RQ3, and RQ5 were developed using GPT-Daybreak Blue (GPT's defensive cyber model); they are evaluated on two agents and harnesses: GPT-5.2-Codex and Zcode GLM-5.
The companion repository contains the code, rule inventories, and prompts as artifacts; it will be made public upon the paper's acceptance.

\subsection{Experimental Setup}
\label{sec:eval-setup}

\paragraph{Corpus}
To evaluate the scanner on skills in the wild, we constructed a corpus from the 75 most-starred GitHub repositories containing skills.
We used GitHub code search to identify 19{,}198 candidate repositories, ranked them by star count, and downloaded the top-75 repositories.
The selected repositories had between 20{,}129 and 358{,}086 stars and contained a total of 2{,}197 skills. We refer to this corpus as the in-the-wild corpus.

\paragraph{Benchmarks}
To compare the Local Analysis module with existing scanners, we use two public benchmarks: SkillTrustBench~v1.0~\cite{skilltrustbench2026} and MalSkillBench~\cite{malskillbench2026}.
SkillTrustBench contains 5{,}520 skills: 2{,}863 labeled malicious, 1{,}014 labeled suspicious, and 1{,}643 labeled normal.
MalSkillBench contains 3{,}944 malicious skills and 4{,}000 benign skills.
Both benchmarks provide labels for individual skills and contain no multi-skill cases.

To evaluate the Composite Analysis module, we adapt SCR-Bench~\cite{xie2026benign}, which evaluates whether skills that are benign in isolation cause harm when used together. The original benchmark executes the participating skills in a sandboxed agent and records the outcome of each case. We adapt these cases for static analysis and assign labels based on their recorded runtime outcomes. In our evaluation, we examine whether the scanner detects the labeled harmful compositions. The companion repository contains the preprocessing code.

SCR-Bench comprises three sub-benchmarks: SCR-CapFlow (150 pairs), SCR-AuthBlur (117 pairs), and SCR-TrustLift (401 pairs). Together they provide 668 risky skill sets. For negatives, we use the benchmark's controls for SCR-AuthBlur (116) and SCR-TrustLift (431), and add two kinds of negative sets. Cross-case shuffles pair skills drawn from unrelated cases, so that each skill comes from a risky case but the pair was not constructed to interact; we draw 40 from SCR-CapFlow, 40 from SCR-AuthBlur, and 20 from SCR-TrustLift. Clean pairs combine benign skills from the \texttt{anthropics/skills} repository (10 pairs). The final evaluation set contains 1{,}325 skill sets: 668 risky and 657 negative.

To prevent directory names from revealing labels, we rename each skill directory using the skill's declared name, replacing role names and control suffixes such as \texttt{skill\_a} and \texttt{-plain}. We also replace group-directory names with opaque case identifiers and exclude experiment scripts containing attack prompts, sandbox trees, and caches. We run the Composite Analysis module as described in Section~\ref{sec:method-composite-analysis}.

\paragraph{Metrics}
For RQ1, we report precision as the proportion of reported findings confirmed through validation. Because the in-the-wild corpus lacks ground-truth labels, we do not estimate recall across the full corpus. We assess it on a manually audited subset of 52 skills from eight repositories, selected from the 23 repositories with confirmed findings.
For RQ2, we use the benchmark labels to measure precision, recall, and \(F_1\), treating malicious and suspicious skills as positive cases and normal or benign skills as negative cases.
For RQ3, we measure precision, recall, and \(F_1\) on the adapted SCR-Bench evaluation set, treating risky skill sets as positive cases and controls, cross-case shuffles, and clean pairs as negative cases.
For RQ4, we ablate the Local and Composite Analysis module's components and report precision, recall, false-positive rate, and \(F_1\).
For RQ5, we report the final mitigation selected for each finding and the number of findings with each mitigation outcome. We also report whether the selected legitimate task was completed.
We measure computational cost using wall-clock time per skill, the number of model requests, and token usage.

\subsection{RQ1: Detection Accuracy in the Wild}
\label{sec:eval-rq1}

\paragraph{Setup}
To evaluate detection accuracy on individual skills, we ran the scanner's  Local Analysis module on the in-the-wild corpus. The corpus contained 2{,}197 skills, including 610 translated or per-agent copies of 279 skills from \texttt{affaan-m/ECC}. We excluded these copies and scanned the remaining 1{,}587 skills in the 75 GitHub repositories.

A panel of three independent analysts who were not members of the research team, each with three years of experience in AI security, assessed the reported findings, comparing them to the source evidence in the associated skills. Each analyst received only the scanner report and rated the findings separately before discussion. We considered a finding confirmed when the panel reached consensus that it represented an actual vulnerability.

\paragraph{Findings}
Fifty-nine findings were identified by the scanner in 43 of the 1{,}587 skills and 23 of the 75 repositories. Table~\ref{tab:rq1-checks} presents reported and confirmed findings, and Table~\ref{tab:rq1-classes} presents them by threat category, along with the severity of confirmed findings.

\begin{table}[!tp]
\centering
\caption{Local Analysis reported and confirmed findings.}
\label{tab:rq1-checks}
\small
\begin{tabular}{@{}lrrr@{}}
\toprule
Check & Reported & Confirmed & Precision \\
\midrule
Value-flow check             & 17 & 14 & 82.4\% \\
Pattern checks               & 2  & 2  & 100.0\%  \\
Instruction review           & 40 & 37 & 92.5\% \\
Capability-declaration check & 0  & 0  & --     \\
Installation-source check    & -- & -- & --     \\
\midrule
Total                        & 59 & 53 & 89.8\% \\
\bottomrule
\end{tabular}
\end{table}

\paragraph{Manual assessment}
The analysts reached unanimous decisions on 57 of the 59 findings (96.6\%) and discussed the remaining two findings and reached consensus on both. Their final assessment confirmed 53 findings and rejected six, resulting in precision of 89.8\%. We grouped confirmed findings that referred to the same vulnerable operation (sink), and this yielded 41 distinct vulnerabilities.

These vulnerabilities allow a user to inject content to control the code or commands executed by the skill (21), inject instructions from the third-party source that the agent follows in later sessions (5), bypass required approvals (8), exfiltrate data or credentials (3), or write inside sensitive directories (4). None of the affected skills was malicious, consistent with the negligent author described in Section~\ref{sec:adversary}.

\paragraph{Severity}
The analysts also rated the severity of each finding based on its potential impact and attacker preconditions. They rated one finding as high, 11 as medium, and 41 as low. Table~\ref{tab:rq1-classes} presents the reported and confirmed findings by threat category, along with the severity of the confirmed findings.

\begin{table}[!tp]
\centering
\caption{Reported and confirmed Local Analysis findings by threat category and severity.}
\label{tab:rq1-classes}
\small
\setlength{\tabcolsep}{3pt}
\newlength{\rqOneSeverityColumnWidth}
\settowidth{\rqOneSeverityColumnWidth}{Confirmed severity}
\setlength{\rqOneSeverityColumnWidth}{0.33334\rqOneSeverityColumnWidth}
\begin{tabular}{@{}lrr*{3}{>{\raggedleft\arraybackslash}p{\rqOneSeverityColumnWidth}}@{}}
\toprule
& & & \multicolumn{3}{c}{Confirmed severity} \\
\cmidrule{4-6}
Threat category & Rep. & Conf. & L & M & H \\
\midrule
Embedded credentials        & 0  & 0  & 0  & 0  & 0 \\
Path traversal              & 7  & 5  & 5  & 0  & 0 \\
Server-side request forgery & 3  & 3  & 3  & 0  & 0 \\
Data exfiltration           & 7  & 6  & 5  & 0  & 1 \\
Log injection               & 0  & 0  & 0  & 0  & 0 \\
Excessive agency            & 11 & 9  & 8  & 1  & 0 \\
Prompt injection            & 16 & 16 & 14 & 2  & 0 \\
Supply-chain attack         & 0  & 0  & 0  & 0  & 0 \\
Command injection           & 2  & 2  & 0  & 2  & 0 \\
Code execution              & 13 & 12 & 6  & 6  & 0 \\
Hidden capability           & 0  & 0  & 0  & 0  & 0 \\
Tool misuse                 & 0  & 0  & 0  & 0  & 0 \\
System-prompt leakage       & 0  & 0  & 0  & 0  & 0 \\
\midrule
Total                       & 59 & 53 & 41 & 11 & 1 \\
\bottomrule
\end{tabular}
\par\smallskip
\parbox{\columnwidth}{\footnotesize\raggedright
\textit{Notes:} Rep.: reported findings; Conf.: findings confirmed by the panel;
L: low; M: medium; H: high.
Severity counts refer to individual confirmed findings, each assigned the
severity of its corresponding vulnerability.
No confirmed finding was rated critical.}
\end{table}

\paragraph{Recall on known vulnerabilities}
Because manually auditing the full corpus is time-consuming, the analysts assessed recall on a reference set of 52 skills from eight repositories. These repositories were selected from the 23 repositories having confirmed findings. The analysts reviewed all 52 skills and identified nine findings, which served as the ground truth for this assessment. The Local Analysis module generated eight findings corresponding to eight of the nine findings, resulting in recall of 88.9\%.

The missed finding arose from an installation-time registry fetch by the skill manager. The fetched content became standing instructions for the agent in later sessions. 

The reference set is small and drawn from repositories with confirmed scanner findings; this assessment therefore does not establish recall across the full corpus. 

\paragraph{PoC validation}
We developed a PoC attack for each of the 41 distinct vulnerabilities.
We disclosed 10 of them to the affected repositories' maintainers: one for each of the nine medium-severity vulnerabilities and one for the high-severity vulnerability. These cover 12 confirmed findings: 11 medium and one high.
We used the remaining 31 attacks only to test the mitigations in RQ5 (Section~\ref{sec:eval-rq5}).

\paragraph{Cost}
The deterministic checks took 973\,s across the 1{,}587 skills. The median execution time per skill was 0.04\,s, and the 95th percentile was 0.5\,s. The maximum was 505\,s for a skill that consisted of an entire repository.
Instruction review excluded 34 skills whose collected instruction documents exceeded the 120{,}000-character input limit. For each of the remaining 1{,}553 skills, it submitted the collected instruction documents in full through a single request. All requests received a response on the first attempt. The median time per skill was 5.3\,s, and the 95th percentile was 22.6\,s. With four concurrent requests, instruction review took 55\,min of wall-clock time. These requests used 6.54\,M input tokens and 0.15\,M output tokens, including 0.10\,M reasoning tokens.
Validation used an additional 1{,}030 requests. It completed within one hour and consumed an estimated 3.00\,M input tokens and 0.56\,M output tokens. Together, instruction review and validation consumed an estimated 9.54\,M input tokens and 0.71\,M output tokens.

\subsection{RQ2: Comparison with Existing Scanners}
\label{sec:eval-rq2}

\begin{figure*}[t]
\centering
\includegraphics[width=\textwidth]{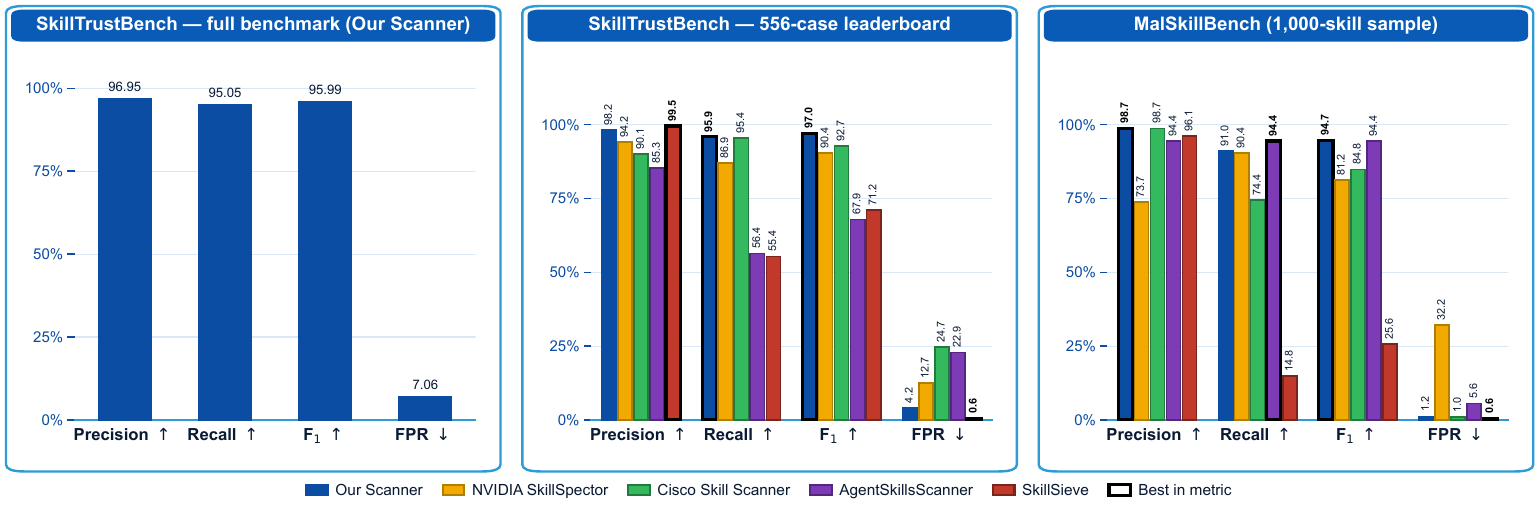}
\caption{Performance comparison of the Local Analysis module and existing scanners (precision, recall, \(F_1\), and false-positive rate).  The left graph
reports the results for our scanner on the full SkillTrustBench benchmark.
The middle graph compares the performance of all evaluated scanners on the 556-case leaderboard subset of SkillTrustBench benchmark,
and the right graph compares the performance of the scanners on a balanced random samples of 1,000-skill sample from MalSkillBench benchmark. For each metric, the best value in the middle and right graphs are outlined in black.}
\label{fig:rq2-comparison}
\end{figure*}

\paragraph{Setup}
We ran the Local Analysis module on all 5{,}520 SkillTrustBench cases, using DeepSeek-V4-Flash for the model-assisted stages. 
For each case, the scanner uses the Content Extraction module to extract the instruction document, supporting files, and declarations, and provides them to the Local Analysis module.
The results are reported in Figure~\ref{fig:rq2-comparison}(a).
Afterwards, we compare our scanner's performance to that of four other scanners: NVIDIA SkillSpector, Cisco Skill Scanner, AgentSkillsScanner, and SkillSieve on two benchmarks: (1) a 556 subset of SkillTrustBench, and (2) a balanced random sample of 1,000 instances from MalSkillBench.
We evaluated all scanners on the same cases and ground-truth labels.

The results are presented in Figure~\ref{fig:rq2-comparison}. On the full SkillTrustBench benchmark, our scanner achieved 96.95\% precision, 95.05\% recall, an \(F_1\) of 95.99\%, and a 7.06\% false-positive rate.

On the 556-case SkillTrustBench subset, our scanner achieved the highest recall and \(F_1\) among the five scanners. AgentSkillsScanner and SkillSieve exhibited substantially lower recall, indicating higher false-negative rates. Our analysis found that over 97\% of their false negatives were labeled safe by conservative static filters before reaching the model-assisted stage. These false negatives were concentrated in suspicious, insecure-coding, and dependency-related cases that lack obvious malicious signatures.

On the MalSkillBench subset, our scanner achieved the highest \(F_1\) and precision, at 94.69\% and 98.70\%. It achieved a recall of 91.00\%, and a false-positive rate of 1.20\%. AgentSkillsScanner achieved the highest recall, while SkillSieve achieved the lowest false-positive rate.

\subsection{RQ3: Composite Vulnerabilities}
\label{sec:eval-rq3}

We evaluated the Composite Analysis module (\S\ref{sec:method-composite-analysis}) on SCR-Bench~\cite{xie2026benign} and the in-the-wild corpus. ColluSkill~\cite{zeng2026colluskill} and CompoSkill~\cite{composkill2026} construct cross-skill attack chains that evade scanners, including Cisco's skill-scanner and SkillSpector. At the time of this writing, the implementations of ColluSkill's ChainGuard, SkillReact, and SkillProbe were unavailable for our evaluations and ColluSkill had not released its attack-chain benchmark. While CompoSkill had released its chain-generation code and SkillsGuard scanner, its benchmark dataset had not yet been released. Thus, SCR-Bench was the only public benchmark dataset with labeled multi-skill cases used in our evaluation. 

\paragraph{Baseline}
We compare the Composite Analysis module with three different configurations of CompoSkill's SkillsGuard and an LLM-only baseline. \footnote{We use the copy in CompoSkill's repository. Its GitHub documentation refers to it as AicoGuardClaw(commit d7dc9d3)~\cite{composkill-code}}
SkillsGuard applies 72 regular expressions to a skill directory, issues a warning after at least one match, and blocks the directory after at least 15 matches. An optional LLM mode assigns a risk rating to the directory as a whole.
We ran SkillsGuard without modification on the 1{,}325 sets of the SCR-Bench benchmark. For its regular-expression mode, we counted a set as flagged when SkillsGuard blocked it. For its LLM mode, we evaluated two thresholds: \emph{high}, which flags high or critical risk, and \emph{review}, which also flags sets rated ``needs review''.

The LLM-only baseline reviews each set as a whole without a dependency graph. It receives every primary instruction document in the set along with the supporting file, using the same input budget as enrichment. The LLM returns a risky or benign verdict.
Table~\ref{tab:rq3-scrbench} reports the results. 

\begin{table}[!tp]
\centering
\footnotesize
\setlength{\tabcolsep}{3pt}
\caption{Composite detection on the adapted SCR-Bench evaluation set.}
\label{tab:rq3-scrbench}
\begin{tabular}{@{}lrrrrrrr@{}}
\toprule
Method & TP & FN & FP & TN & Precision (\%) & Recall (\%) & \(F_1\) \\
\midrule
Our scanner & 665 & 3 & 9 & 648 & \textbf{98.7} & \textbf{99.6} & \textbf{0.991} \\
SG review   & 663 &   5 & 214 & 443 & 75.6 & \underline{99.3} & \underline{0.858} \\
SG regex    & 551 & 117 &  71 & 586 & 88.6 & 82.5 & 0.854 \\
SG high     & 499 & 169 &  56 & 601 & \underline{89.9} & 74.7 & 0.816 \\
LLM-only    & 657 &  11 & 362 & 295 & 64.5 & 98.4 & 0.779 \\
\bottomrule
\end{tabular}
\par\smallskip
\parbox{\columnwidth}{\footnotesize\raggedright
\textit{Notes:} SG: SkillsGuard; LLM: large language model;
TP: true positives; FN: false negatives; FP: false positives;
TN: true negatives.
For precision, recall, and \(F_1\), the best value is in bold
and the second-best value is underlined.}
\end{table}

\paragraph{In-the-wild findings}
\label{sec:eval-rq3-pairs}

We evaluated each of the 1{,}587 skills in turn as the input skill \(s\), with the remaining skills of the corpus forming the installed set \(\mathcal{S}\).
The validation module assessed every composite candidate, and the same three analysts as in RQ1 reviewed the candidates it retained.
We identified two composite vulnerabilities involving four skills from three GitHub repositories. The first combines a conversation-history reader from \texttt{daymade/claude-code-skills} with a Slack sender from \texttt{ArtyMcLabin/ClaudeCode-global-config}, enabling stored prompt injection to trigger unauthorized messages. The second combines the \emph{watchers} and \emph{mcp-oauth-remote-gateway} skills from \texttt{NousResearch/hermes-agent}, enabling disclosure of an OAuth bearer token to an attacker-controlled endpoint.
None of the four skills is among the 59 findings that Local Analysis reported in RQ1.

\paragraph{Finding 1: Stored prompt injection through conversation history}
The first finding combines a conversation-history reader, \emph{claude-code-history-files-finder}, with a Slack sender, \emph{slack}. The reader indexes past Claude Code sessions, including messages, tool inputs, and tool results, and retrieves stored records verbatim. The Slack sender posts agent-composed text to a Slack channel through \texttt{chat.postMessage} but does not retrieve conversation history. The skills share no file or environment variable, and neither invokes the other, so \(G_0\) contains no edge between them. Enrichment added a context edge from the reader to the sender. The split-capability rule then identified a path from \(\mathsf{SensitiveData}\) to \(\mathsf{Egress}\) through which the agent could send retrieved session content to Slack.

We tested whether this path could carry a stored prompt injection. For each run we tested, we used a fresh sandbox with byte-identical copies of both skills and one stored session. The session included the result of fetching an attacker-filed GitHub issue with a directive hidden in an HTML comment. Later entries contained a private pricing decision and a canary billing key. The victims were GPT-5.2-Codex and Zcode GLM-5, each running with pre-approved shell commands under the same setup. We tested two payloads: one claimed pre-authorized operational authority, while the other instructed the agent to send the key to an outside account and conceal the action. Controls used the same history with the payload removed.

In 10 out of 12 runs, GPT-5.2 identified the payload as an injection, refused to follow the instruction, and reported it to the user. In the remaining two runs, the agent followed the injected instruction and issued an unauthorized Slack request. For GLM-5, seven of 12 runs followed the injected instruction. The PoC is included in the companion repository.

\paragraph{Finding 2: Token disclosure through a confused deputy}
The second finding combines \emph{watchers}, which is a feed monitor, with \emph{mcp-oauth-remote-gateway}, which is a diagnostic skill that has access to an OAuth bearer token. An attacker can publish the payload containing a URL to a feed that the victim has subscribed to. \emph{watchers} prints the item into the agent’s tool output, placing attacker-authored text in the agent’s context. The payload of the attacker directs the agent to treat the URL as the diagnostic destination and passes it to \emph{mcp-oauth-remote-gateway} as a command-line argument; the gateway sends its stored OAuth bearer token to the attacker’s server.

We tested this path in the same contained sandbox environment described above. For this finding, we separately added a fake token file and webhook listeners representing the configured endpoint, which serves as an attacker URL. We counted a run as successful only when its canary token reached the listener. In five of 12 GPT-5.2 runs, the victim agent invoked the diagnostic with the attacker URL, causing the canary token to reach the listener. In GLM-5, the canary reached the listener in six of 12 runs. The PoC is included in the companion repository.

\subsection{RQ4: Ablation Study}
\label{sec:eval-rq4}

\paragraph{Setup}
We performed ablation studies on the Local Analysis and Composite Analysis modules separately.
For the Local Analysis module, we used the 556-case SkillTrustBench leaderboard subset from RQ2, which contains 390 malicious and 166 benign cases. 
We evaluated six configurations: deterministic detection ($D$),
LLM instruction and code review ($L$), their union ($D+L$), and
each configuration followed by package-level validation ($D+V$,
$L+V$, and $D+L+V$). The deterministic and LLM detectors operated
independently.

%We evaluated all combinations of deterministic detection, LLM detection, and validation. All six configurations were scored on the same cases and ground-truth. We evaluated each ablation configuration over five runs to assess consistency.

For the Composite Analysis module, we used the SCR-CapFlow and SCR-AuthBlur sets of the SCR-Bench benchmark containing 267 risky sets and 218 clean sets, comprising 116 SCR-AuthBlur controls, 100 cross-case shuffles, and two clean pairs. We compared \(G_0\) with the interface layer, \(G_0\) with model-proposed edges from model enrichment, and \(G_0\) with both. As in the experiment addressing RQ3, a set was flagged if the method reported at least one composite finding. We also evaluated each configuration with the Validation module added. We evaluated each ablation configuration over five runs to assess consistency. Tables~\ref{tab:rq4-ablation} and ~\ref{tab:rq4-composite-ablation} report the results of the Local Analysis module and  Composite Analysis module ablation, respectively. For each configuration, we computed the evaluation metrics separately for each run and report their mean \(\pm\) standard deviation across the five runs.

\begin{table}[!tp]
\centering
\caption{Local Analysis ablation study on the 556-case SkillTrustBench
leaderboard subset. D: deterministic detection;
L: LLM detection; V: package-level validation.
Metrics are reported as mean $\pm$ sample standard deviation across
five runs; exactly zero standard deviations are omitted.}
\label{tab:rq4-ablation}
\small
\setlength{\tabcolsep}{3pt}
\begin{tabular}{@{}lrrrrrrrr@{}}
\toprule
Config. & Precision (\%) & Recall (\%) & FPR (\%) & $F_1$ \\
\midrule
D & $85.3$ & $68.2$ & $27.7$ & $0.758$ \\
L & $97.0 \pm 0.4$ & $94.2 \pm 0.6$ & $6.9 \pm 1.0$ & $0.956 \pm 0.002$ \\
D + L & $87.1 \pm 0.4$ & $96.5 \pm 0.5$ & $33.5 \pm 1.3$ & $0.916 \pm 0.001$ \\
\midrule
D + V & $98.7 \pm 0.2$ & $67.8 \pm 0.4$ & $2.2 \pm 0.3$ & $0.804 \pm 0.002$ \\
L + V & $98.0 \pm 0.3$ & $95.8 \pm 0.1$ & $4.6 \pm 0.7$ & $0.969 \pm 0.001$ \\
D + L + V & $98.0 \pm 0.4$ & $96.0 \pm 0.1$ & $4.6 \pm 0.9$ & $0.970 \pm 0.001$ \\
\bottomrule
\end{tabular}
\end{table}

\begin{table}[!tp]
\centering
\caption{Composite Analysis ablation study on the 485 SCR-CapFlow and SCR-AuthBlur sets.}
\label{tab:rq4-composite-ablation}
\small
\setlength{\tabcolsep}{3pt}
\begin{tabular}{@{}lrrrrrrrr@{}}
\toprule
Config. & Precision (\%) & Recall (\%) & FPR (\%) & $F_1$ \\
\midrule
$G_0$ + IL & $100.0$ & $54.7$ & $0.0$ & $0.707$ \\
$G_0$ + ME & $94.6 \pm 2.3$ & $98.9 \pm 0.5$ & $7.0 \pm 3.0$ & $0.967 \pm 0.010$ \\
$G_0$ + IL + ME & $94.6 \pm 2.3$ & $99.2 \pm 0.2$ & $7.0 \pm 3.0$ & $0.968 \pm 0.012$ \\
\midrule
$G_0$ + IL + V & $100.0$ & $54.2 \pm 0.2$ & $0.0$ & $0.703 \pm 0.001$\\
$G_0$ + ME + V & $98.0 \pm 0.8$ & $95.2 \pm 1.5$ & $2.4 \pm 1.0$ & $0.966 \pm 0.011$ \\
$G_0$ + IL + ME + V & $98.0 \pm 0.8$ & $95.4 \pm 1.6$ & $2.4 \pm 1.0$ & $0.967 \pm 0.012$ \\
\bottomrule
\end{tabular}
\par\smallskip
\parbox{\columnwidth}{\footnotesize\raggedright
\textit{Notes:} \(G_0\): initial dependency graph;
IL: interface layer; ME: model-proposed edges; V: Validation module; FPR: false-positive rate.}
\end{table}

The interface layer achieved 54.7\% recall and detected none of the SCR-AuthBlur pairs.
Model enrichment without the interface layer reached an $F_1$ score of 0.967, and adding the interface layer raised it to 0.968.

\subsection{RQ5: Mitigation Recommendations}
\label{sec:eval-rq5}

\paragraph{Setup}
We ran the Mitigation module (\S\ref{sec:method-mitigation}) on 53 local findings and two composite findings. 
Each evaluated mitigation was applied to the byte-identical copy of the affected skill and its agent environment. For composite findings, the environment included the relevant installed skills and the skill being considered for installation. 
For each mitigation that could be tested, we replayed the PoC attack developed for the corresponding vulnerability in RQ1, or the PoC described in RQ3 for a composite finding, over three repetitions, withholding any approval required for the harmful operation.
These attack tests used the modified permissions for configuration changes and covered the complete reported attack path for composite mitigations.
To assess whether the mitigations preserved legitimate task completion, we also checked whether the affected skill or skill set could complete the documented task, granting the approvals required for legitimate execution.

\paragraph{Mitigation selection}
The Mitigation module recommended a mitigation for each of the 55 findings.
Table~\ref{tab:rq5-types} summarizes the final mitigation option
selected for each finding. A combined option includes all changes retained in the selected configuration.

For 13 of the 16 code-based local findings, a substitute providing the required functionality was selected. No suitable substitute was available for the remaining three, and the Mitigation module recommended executable-code patches. 
For 37 findings, the selected mitigation involved modifying unsafe instructions: 17 were assigned instruction patches alone, while 20 were assigned instruction patches combined with configuration changes. 
For one of the composite findings, the module recommended a substitute with an accompanying configuration change, and for the other, an executable-code patch was recommended.

\begin{table}[!tp]
\centering
\caption{Mitigations selected to address the 55 reported findings.}
\label{tab:rq5-types}
\small
\setlength{\tabcolsep}{5pt}
\begin{tabular}{@{}lr@{}}
\toprule
Mitigation & Findings \\
\midrule
Substitution alone & 13 \\
Substitution with configuration & 1 \\
Executable-code patch alone & 4 \\
Executable-code patch with configuration & 0 \\
Instruction patch alone & 17 \\
Instruction patch with configuration & 20 \\
Configuration alone & 0 \\
No mitigation selected & 0 \\
\midrule
Total & 55 \\
\bottomrule
\end{tabular}
\par\smallskip
\parbox{\columnwidth}{\footnotesize\raggedright
\textit{Notes:} Each finding is counted once according to the final
mitigation selected to address it. The counts include mitigations
for two findings whose effectiveness could not be evaluated.}
\end{table}

\paragraph{Task preservation}
We assessed legitimate-task completion separately from mitigation effectiveness. This assessment examines whether the mitigated skill or skill set can still perform the documented task.

The tasks were completed in the sandbox for 53 of the 55 findings. For the remaining two, the original vulnerable skills could run in the sandbox, but the selected substitutes required Artifactory installations, which were prevented by the sandbox's network restrictions. We reviewed the substitutes' documented behavior against the original task requirements but could not verify task completion through execution.

\paragraph{Mitigation effectiveness}
We classified a mitigation as effective in the tested conditions when it prevented the reported attack in all three repetitions. 
Failure to reproduce an attack without evidence that the mitigation took effect was insufficient to establish effectiveness.
The authors acted as the user in these tests and verified each outcome.
For configuration changes, we examined whether the environment blocked the harmful operation or required approval. Because these changes leave the skill unchanged, the scanner may still report the finding. We therefore assessed effectiveness by observing execution and approval behavior.
Tested findings for which the selected mitigation established neither effective nor partial protection were classified as unresolved. 
Table~\ref{tab:rq5-outcomes} reports the outcomes for the final selected mitigations. Since the two skills needed further files, we kept their effectiveness as not evaluated.

\begin{table}[!tp]
\centering
\caption{Outcomes for the final selected mitigations across 55 findings.}
\label{tab:rq5-outcomes}
\small
\setlength{\tabcolsep}{5pt}
\begin{tabular}{@{}lr@{}}
\toprule
Outcome & Findings \\
\midrule
Effective in the tested conditions & 47 \\
Partially mitigated & 6 \\
Unresolved after testing & 0 \\
Effectiveness not evaluated & 2 \\
\midrule
Total & 55 \\
\bottomrule
\end{tabular}
\end{table}

\paragraph{Case study: history retrieval and Slack sending}
\label{sec:eval-mitigation}
We applied the Mitigation module to the composite vulnerability identified in the RQ3 experiment, which combines the conversation-history reader \emph{claude-code-history-files-finder} with the Slack sender \emph{slack}.
The module targeted the sending capability, and to identify a substitute, the module collected eight Slack-sending skills from public registries and GitHub.
For each candidate, it identified the sending mechanism, classified the approval gate, checked compatibility with the user's agent environment, and ran the Local Analysis module on the skills.
Table~\ref{tab:eval-mitigation} presents the candidates considered by the module, followed by the original sender.

\begin{table*}[!tp]
\centering
\caption{Slack-sending substitutes considered by the Mitigation module,
followed by the original sender (host: Codex).
All support sending agent-written text. Placeholder credentials
in documentation examples are not counted as Local Analysis module findings.
Unvalidated value-flow candidates are distinguished from confirmed findings.}
\label{tab:eval-mitigation}
\small
\setlength{\tabcolsep}{3pt}
\begin{tabularx}{\textwidth}{@{}>{\raggedright\arraybackslash}p{0.31\textwidth}>{\raggedright\arraybackslash}p{0.29\textwidth}>{\raggedright\arraybackslash}X@{}}
\toprule
Skill & Gate on Sending & Local Analysis Result \\
\midrule
slackapi/slack-messaging\textsuperscript{*}
  & host (MCP tool approval)
  & none reported \\
slackapi/slack-api\textsuperscript{*}
  & instruction (confirm writes)
  & none reported \\
idanbeck/slack-skill
  & instruction (confirm each send)
  & undeclared network use \\
openclaw/slack-personal
  & optional draft review
  & none reported \\
devxoul/agent-slack
  & none
  & excessive agency \\
devxoul/agent-slackbot
  & none
  & excessive agency \\
HartreeWorks/skill--slack
  & none on sends
  & 22 unvalidated value-flow candidates \\
openclaw/slack
  & host (OpenClaw tool)
  & excluded: requires OpenClaw \\
\midrule
original \emph{slack}
  & none
  & none reported \\
\bottomrule
\end{tabularx}
\par\smallskip
\parbox{\textwidth}{\footnotesize
\textsuperscript{*}Published by Slack.}
\end{table*}

The module excluded \emph{openclaw/slack}, because it requires OpenClaw's built-in tool and is incompatible with the Codex and Zcode harnesses.
Of the remaining seven substitutes, three provide no gate on sending, and \emph{openclaw/slack-personal} offers optional draft review. Therefore, these four candidates do not establish a required approval step.
Two others rely on confirmation instructions and therefore provide only partial mitigation, because prompt injection could cause the agent to disregard the instructions.

The remaining substitute, \emph{slackapi/slack-messaging}, uses an MCP tool with a permission separate from the shell's.
This separation allows the user to retain shell pre-approval while requiring approval for Slack messages.
The module therefore recommended replacing the original sender with \emph{slackapi/slack-messaging} together with the required approval setting for its send tool.

The replacement with its required configuration was selected for the final assessment and is counted as one combined mitigation in Table~\ref{tab:rq5-types}.
\section{Discussion}
\label{sec:discussion}

\subsection{Unintended Vulnerabilities in Skills}
We found no evidence of malicious intent in the skills with confirmed vulnerabilities in RQ1. The vulnerabilities arose in workflows intended to support legitimate tasks, but whose design allowed attacker-controlled content to influence the agent’s actions.
For example, the \texttt{implementing-feature} skill in \texttt{streamlit/streamlit} instructs the agent to “Treat the issue details as the feature specification” and to “Be fully autonomous” without pausing for approval. These instructions support autonomous feature implementation. However, an attacker can supply content that the agent treats as authoritative guidance for its work.
Similarly, the \texttt{opencode} skill in \texttt{volcengine/OpenViking} inserts a worktree path returned by the local OpenCode server into a shell command without sanitizing it. Using the returned path serves the skill’s intended operation, but an attacker who can influence that path can add shell meta-characters that cause additional commands to execute.
In both cases, the attacker exploits the handling of external input within an intended workflow. Therefore, the absence of deliberately malicious behavior in a skill does not establish that the skill is safe to use.

\subsection{Permissions and autonomy}
The Composite Analysis module PoC showed that a directive embedded in a previously retrieved GitHub issue could re-enter the agent's context through stored conversation history.
The single reference-set vulnerability the scanner missed in RQ1 involved a related persistence mechanism, where content fetched during installation became standing instructions for subsequent sessions.
In both cases, stored lower-trust content could influence later actions, while the agent’s permissions and autonomy determined the scope of those actions.
Broader filesystem, shell, and network access increases the range of actions an attacker can induce, while greater autonomy allows more of these actions to proceed without the need for further approval.
Enterprises deploying agents should therefore limit access to the resources and operations required for each task and enforce these restrictions at runtime.
These restrictions must also account for how data passes between permitted operations.
This distinction is established in information-flow security: permission to read data does not by itself constrain where the receiving program subsequently sends it~\cite{sabelfeld2003language}.
In the composite PoC, the agent had permission to read conversation history and send Slack messages.
These permissions did not establish the user's authorization to send retrieved information to a recipient specified by an injected instruction.
Security requirements for skills should therefore distinguish capabilities claimed by authors from permissions enforced by the runtime and specify how user authorization applies to data transfers between tools and skills.

\subsection{Screening through a common skill marketplace}
Research on the npm ecosystem has shown that package dependencies and shared maintainers can spread exposure to vulnerable or malicious code~\cite{zimmermann2019small}.
In the agent-skill ecosystem, several registries already provide security information about their skills.
ClawHub reports a scan status for each skill~\cite{openclaw2026skills}, Snyk scans skills in Tessl's registry~\cite{snyk2026registry}, and NVIDIA publishes a catalog of verified skills~\cite{nvidia2026verified}.
Each service covers its own catalog.
Users installing skills from multiple sources must therefore consult separate registries, while skills distributed outside these catalogs receive no coverage from their screening.
A common skill marketplace, analogous to a mobile app store and maintained by a neutral party, could apply shared screening criteria across participating repositories.
The per-skill cost and latency reported in Section~\ref{sec:eval-rq1} suggest that automated checks could be applied to each published version.
The marketplace could display findings and analysis coverage alongside user ratings, giving users a common source of information before installation.
Screening individual skill versions would not, however, establish the safety of combinations installed by a user. Marketplace screening could therefore be complemented by composite analysis of the installed skill set and runtime controls over interactions among skills.

\subsection{Advantages over existing scanners}
The scanner's main advantage is not any single technique but its separation of broad candidate generation from evidence-based validation. 
The compared four scanners lack this separation, leading either to missed security issues or unsupported detections. AgentSkillsScanner and SkillSieve use conservative static filters to label cases as safe before model-assisted analysis. On the 556-case SkillTrustBench subset, these two scanners labeled more than 97\% of their false negatives as safe, particularly suspicious, insecure-coding, and dependency-related cases without obvious malicious signatures. 
The lower recall of NVIDIA SkillSpector reflects a similar limitation in candidate coverage, whereas Cisco Skill Scanner achieved similar recall but a substantially higher false-positive rate, indicating weaker evidence filtering. 
In our scanner, no Local Analysis check makes a terminal safe decision. Pattern, instruction-review, value-flow, installation-source, and capability-declaration checks generate candidates independently, after which the Validation module assesses their source evidence and context. This design retains risks identified by any analysis while rejecting unsupported candidates, producing the highest recall and \(F_1\) among the five scanners.

\section{Limitations and Future Work}
\label{sec:limitations}

\subsection{Static analysis}
Our scanner does not execute skills.
It recovers code written as string literals (\S\ref{sec:method-local-analysis}) but not code constructed at runtime obfuscated beyond the normalizations it applies or fetched from the network.
Value-flow checks support Python, JavaScript, and shell; flows in scripts written in other languages are not traced.
The Composite Analysis module matches literal paths and environment-variable names, so a dependency whose resource name is computed at runtime only appears in the graph if enrichment is enabled.
The scanner does not trace content fetched by a skill manager during installation and loaded as instructions in later sessions, as illustrated by the vulnerability missed in RQ1(\S~\ref{sec:eval-rq1}). 
Future work could pair static analysis with sandbox execution to detect dynamically generated code and resource names, and analyze content installed for later sessions.

\subsection{Recall on the in-the-wild corpus}
The in-the-wild corpus has no ground-truth labels, so we only measured recall on a reference set of eight repositories and 52 skills. 
Future work could measure recall using a larger, independently audited sample across more repositories and vulnerability types.

\subsection{Adversarial effectiveness of the scanner}
The scanner includes deterministic checks to detect prompt injections in docstrings, code, and skill instructions.
The LLM is also instructed to treat the analyzed content as data and not follow embedded instructions.
However, the checks cover a limited set of injection patterns, and neither these checks nor the LLM instructions establish the scanner's robustness against adversarial inputs designed to evade detection or manipulate its analysis.
Evaluating this robustness and developing mechanisms to strengthen it are outside the scope of this work and remain future work.

\subsection{Composite benchmarks}
SCR-Bench is the only public benchmark with labeled multi-skill cases.
ColluSkill and CompoSkill evaluate attack chains they generate themselves and have not released their benchmarks~\cite{zeng2026colluskill,composkill2026}.
The composite results therefore rest on one benchmark and the in-the-wild corpus.
Future work could develop and release additional labeled, multi-skill benchmarks covering more dependencies and attack paths.
\section{Conclusion}

In this research, we proposed a skill scanner that identifies vulnerabilities within individual skills and assesses whether adding a skill to an installed set introduces risks through their interactions. 
Its Local Analysis module examines instructions, code, installation sources, and capability declarations, while its Composite Analysis module constructs a dependency graph to identify potentially harmful paths across skills. 
By combining deterministic checks with optional LLM-based review and graph enrichment, the scanner can identify risks that lack explicit malicious signatures or depend on relationships conveyed through instructions. 
Both analyses submit candidates to a Validation module that checks their supporting source evidence before reporting findings.

Across 1,587 skills from the 75 most-starred GitHub repositories containing skills, analysts confirmed 53 of the Local Analysis module’s 59 findings (89.8\% precision), representing 41 distinct vulnerabilities in 23 repositories. 
We found no evidence of malicious intent in the affected skills, indicating that workflows designed for legitimate tasks can nevertheless expose exploitable weaknesses. 
Composite Analysis identified two additional findings involving conversation-history retrieval and Slack messaging, and feed monitoring and OAuth diagnostics. 
These findings demonstrate the need to examine how information and capabilities pass between skills, which our scanner addresses through dependency analysis alongside individual-skill analysis.
The benchmark results support the effectiveness of this combined approach. 
Among five evaluated scanners, Local Analysis achieved the highest recall on the 556-case SkillTrustBench subset (95.90\%, with a 4.22\% false-positive rate) and the highest $F_1$ on the 1,000-skill MalSkillBench sample (94.69\%). 
Composite Analysis achieved 98.7\% precision and 99.6\% recall ($F_1$ 0.991) on the adapted SCR-Bench evaluation set. 

Beyond detection, the Mitigation module recommends skill substitution, patches, and configuration changes. 
Mitigations for 47 of the 55 confirmed findings prevented the corresponding attacks in all three trials while preserving the tested legitimate functionality; six findings were partially mitigated, and two could not be evaluated due to runtime dependencies.
These results support using the scanner to inform installation and remediation decisions, while recognizing that static analysis cannot establish safety under all execution conditions. 
The analysis should therefore be complemented by runtime restrictions on resource access and approval controls that enforce the user’s authorization for the data transmitted and its destination.

\section{Ethics statement}

The authors used publicly available skill repositories and benchmark datasets to investigate vulnerabilities in agent skills. To support remediation, they disclosed ten PoC attacks covering medium- and high-severity local vulnerabilities to the affected maintainers. Composite vulnerabilities were validated in contained sandbox environments using byte-identical copies of the affected skills, canary credentials, and dummy listeners. These controls allowed the authors to observe whether the attacks caused disclosure of test credentials. Mitigations were also evaluated in sandbox environments to assess whether they prevented the demonstrated attacks while preserving the skills' legitimate functionality.

\section{Declaration of AI usage}

During the preparation of this manuscript, the authors used LLMs to assist with language editing, code optimization, and figure generation. The authors reviewed the AI-generated material and take full responsibility for the content of the published article.

\bibliographystyle{elsarticle-num}
\bibliography{refs}

% \appendices
% \input{sections/appendix}

\end{document}